\documentclass[%
 aip,
 amsmath,amssymb,
 preprint,%
]{revtex4-1}

\draft % marks overfull lines with a black rule on the right

\usepackage{hyperref}
\hypersetup{colorlinks,allcolors=black}

\usepackage{graphicx}% Include figure files
\usepackage{dcolumn}% Align table columns on decimal point
\usepackage{bm}% bold math
\usepackage[final]{changes} %Manually Track changes - turned off

\usepackage[utf8]{inputenc}
\usepackage[T1]{fontenc}
\usepackage{mathptmx}
\usepackage{etoolbox}
\usepackage{units}
\usepackage{tikz}
\definecolor{myyellow}{rgb}{0.9290,0.6940,0.1250}
\definecolor{myred}{rgb}{0.8500,0.3250,0.0980}
\definecolor{myblue}{rgb}{0,0.3352,0.5557}
\usetikzlibrary{decorations.pathreplacing,angles,quotes}
\usetikzlibrary{arrows}
\tikzset{
  double -latex/.style args={#1 colored by #2 and #3}{    
    -latex,line width=#1,#2,
    postaction={draw,-latex,#3,line width=(#1)/3,shorten <=(#1)/4,shorten >=4.5*(#1)/3},
  },
  double round cap-latex/.style args={#1 colored by #2 and #3}{    
    round cap-latex,line width=#1,#2,
    postaction={draw,round cap-latex,#3,line width=(#1)/3,shorten <=(#1)/4,shorten >=4.5*(#1)/3},
  },
  double round cap-stealth/.style args={#1 colored by #2 and #3}{
    round cap-stealth,line width=#1,#2,
    postaction={round cap-stealth,draw,,#3,line width=(#1)/3,shorten <=(#1)/3,shorten >=2*(#1)/3},
  },
  double -stealth/.style args={#1 colored by #2 and #3}{
    -stealth,line width=#1,#2,
    postaction={-stealth,draw,,#3,line width=(#1)/3,shorten <=(#1)/3,shorten >=2*(#1)/3},
  },
}

\makeatletter
\def\@email#1#2{%
 \endgroup
 \patchcmd{\titleblock@produce}
  {\frontmatter@RRAPformat}
  {\frontmatter@RRAPformat{\produce@RRAP{*#1\href{mailto:#2}{#2}}}\frontmatter@RRAPformat}
  {}{}
}%
\makeatother
\begin{document}

\preprint{POF22-AR-01287}

\title[The Stability of Wakes of Floating Wind Turbines]{The Stability of Wakes of Floating Wind Turbines}
% Force line breaks with \\
\author{V. G. Kleine}
  \email{vitok@mech.kth.se}
  \affiliation{FLOW Turbulence Lab., Dept. of Engineering Mechanics, KTH Royal Institute of Technology, SE-100 44 Stockholm, Sweden}%
  \affiliation{Division of Aeronautical and Aerospace Engineering, Instituto Tecnol\'{o}gico de Aeron\'{a}utica, 12228-900, S\~{a}o Jos\'{e} dos Campos - SP, Brazil}%

\author{L. Franceschini}
  \affiliation{USP University of S\~{a}o Paulo, Escola Polit\'{e}cnica, Department of Mechanical Engineering, Av. Prof. Mello Moraes, 2231, Cidade Universit\'{a}ria, 05508-030, S\~{a}o Paulo – SP, Brazil}%
  \affiliation{Physics of Fluids Group, Max Planck Center Twente for Complex Fluid Dynamics, University of Twente, 7500 AE Enschede, The Netherlands}

\author{B. S. Carmo}
  \affiliation{USP University of S\~{a}o Paulo, Escola Polit\'{e}cnica, Department of Mechanical Engineering, Av. Prof. Mello Moraes, 2231, Cidade Universit\'{a}ria, 05508-030, S\~{a}o Paulo – SP, Brazil}%

\author{A. Hanifi}
\author{D. S. Henningson}
  \affiliation{FLOW Turbulence Lab., Dept. of Engineering Mechanics, KTH Royal Institute of Technology, SE-100 44 Stockholm, Sweden}%

\date{\today}% It is always \today, today,
             %  but any date may be explicitly specified

%\begin{document}

\title{The Stability of Wakes of Floating Wind Turbines}

% \Author[affil]{given_name}{surname}

%\Author[1,2]{Vitor G}{Kleine}
%\Author[3]{Lucas}{Franceschini}
%\Author[3]{Bruno Souza}{Carmo}
%\Author[1]{Ardeshir}{Hanifi}
%\Author[1]{Dan S}{Henningson}

%\affil[1]{FLOW Turbulence Lab., Dept. of Engineering Mechanics, KTH Royal Institute of Technology, SE-100 44 Stockholm, Sweden}
%\affil[2]{Instituto Tecnol\'{o}gico de Aeron\'{a}utica, Pra\c{c}a Marechal Eduardo Gomes, 50, Vila das Ac\'{a}cias, 12228-900, S\~{a}o Jos\'{e} dos Campos - SP, Brazil}
%\affil[3]{USP University of S\~{a}o Paulo, Escola Polit\'{e}cnica, Mechanical Engineering Department, Av. Prof. Mello Moraes, 2231, Cidade Universit\'{a}ria, 05508-030, S\~{a}o Paulo – SP, Brazil}

%% The [] brackets identify the author with the corresponding affiliation. 1, 2, 3, etc. should be inserted.

%% If an author is deceased, please mark the respective author name(s) with a dagger, e.g. "\Author[2,$\dag$]{Anton}{Smith}", and add a further "\affil[$\dag$]{deceased, 1 July 2019}".

%% If authors contributed equally, please mark the respective author names with an asterisk, e.g. "\Author[2,*]{Anton}{Smith}" and "\Author[3,*]{Bradley}{Miller}" and add a further affiliation: "\affil[*]{These authors contributed equally to this work.}".

%\correspondence{Vitor Kleine (vitok@mech.kth.se)}

%\runningtitle{The Stability of Wakes of Floating Wind Turbines}

%\runningauthor{Kleine et al.}

%\received{}
%\pubdiscuss{} %% only important for two-stage journals
%\revised{}
%\accepted{}
%\published{}

%% These dates will be inserted by Copernicus Publications during the typesetting process.

%\firstpage{1}

\begin{abstract}
Floating offshore wind turbines (FOWTs) are subjected to platform motion induced by wind and wave loads. The oscillatory movement trigger vortex instabilities, modifying the wake structure, influencing the flow reaching downstream wind turbines. In this work, the wake of a FOWT is analysed by means of numerical simulations and comparison with linear stability theory. Two simplified models based on the stability of vortices are developed for all degrees of freedom of turbine motion. In our numerical simulations, the wind turbine blades are modeled as actuator lines and a spectral-element method with low dispersion and dissipation is employed to study the evolution of the perturbations. The turbine motion excites vortex instability modes predicted by the linear stability of helical vortices. The flow structures that are formed in the non-linear regime are a consequence of the growth of these modes and preserve some of the characteristics that can be explained and predicted by the linear theory. The number of vortices that interact and the growth rate of disturbances are well predicted by a simple stability model of a two-dimensional row of vortices. For all types of motion, the highest growth rate is observed when the frequency of motion is one and a half the frequency of rotation of the turbine; that induces the out-of-phase vortex pairing mechanism. For lower frequencies of motion, several vortices coalesce to form large flow structures, which cause high amplitude of oscillations in the streamwise velocities, that may increase fatigue or induce high amplitude motion on downstream turbines.
\end{abstract}

\maketitle

%\copyrightstatement{TEXT} %% This section is optional and can be used for copyright transfers.

\section{Introduction} \label{sec:introduction}

Offshore wind energy has been receiving an increasing amount of attention in the past few years. The reason for this is the several advantages it has in comparison with onshore wind turbines, such as: stronger and more consistent wind conditions, the possibility to install larger turbines, the reduced impact of the noise generated by the turbines and the mitigated harm it causes to wildlife. In order to make this kind of energy accessible and more affordable, a few technological issues need to be overcome. For example, for deep waters (typically above 60 meters), the bottom-mounted configuration is no longer feasible and Floating Offshore Wind Turbines (FOWT) have to be considered, where, inspired by naval architecture solutions, the turbine is usually mounted on a floating platform.

Since this platform is not bottom-mounted, it is allowed (under the constraint of the mooring system) to move under the action of the wind, sea waves and sea currents. This movement may produce modifications on the structure of the wake of the turbine, possibly having important performance impacts in other downstream turbines. Some studies have been carried out in the past in that direction. For example, the performance of the wind turbine (such as power and thrust) was evaluated using imposed pitch \citep{tran2015platform,fang2020numerical} and surge \citep{wen2017influences} motion. Later, other studies \citep{tran2016fully,tran2016cfd} performed numerical simulations in which the platform motion was not imposed but computed. More recently, the effect of the motion of turbines mounted on spar and submersible platforms on the generation of power was assessed through numerical simulations coupling the blade element momentum to calculate the blade loads, a control model for the turbine, a simplified hydrodynamics model to account for the platform motion, and LES (Large Eddy Simulation) of the free wind \citep{Johlas2021WE}. However, all those studies have paid little attention to the far-wake generated by the turbine, focusing only on near-turbine phenomena affecting the performance of that turbine. The predominant numerical model employed in those works was URANS (Unsteady Reynolds-Averaged Navier-Stokes), in which the flow close to \added{the} blades was resolved.

Vortex instabilities in the near wake of FOWTs were already observed by \citet{sebastian2012analysis} and \citet{farrugia2016study}. Recently, a few studies \citep{lee2019effects,dong2019modified} have paid more attention to the issue of the far wake when platform movement occurs. To avoid the dissipation of traditional low-order computational fluid dynamics (CFD) methods, the wake was modelled by Lagrangian-based vortical particles \citep{lee2019effects}, vortex rings \citep{dong2019modified} or a free vortex wake method \citep{sebastian2012analysis,farrugia2016study}. With these vortex-based models, they were able to provide evidence of possible patterns for the interaction of tip-vortices when the platform undergoes motion on each of its six solid degrees of freedom. However, their work focused more on the forces and performance of the turbine than on the wake, so the stability of the wake was not thoroughly explored.

Alternatively, several numerical studies have been carried out focusing on the stability of tip vortices of fixed wind turbines. Those studies are important for understanding how the wakes of several wind turbines interact with each other in a wind farm and how it affects its global performance. Earlier studies \citep{widnall1972stability,gupta1974theoretical} developed a theoretical framework to study the stability of helical vortices. The mutual inductance mode predicted by \citet{widnall1972stability} has been observed in experiments with models of ship propellers \citep{felli2011mechanisms} and wind turbines \citep{sherry2013interaction} as the main mechanism of vortex interaction. The results of the stability theory have been confirmed quantitatively both experimentally  \citep{leweke2014long,nemes2015mutual,quaranta2015long,quaranta2019local} and numerically \citep{ivanell2010stability,sarmast2014mutual}.  Later, those instabilities were studied numerically on a flow under different configurations such as shear \citep{kleusberg2019tip} and yaw misalignment \citep{kleusberg2019wind}. More recently the interaction of vortices of two in-line turbines was studied \citep{kleine2019tip}. These recent numerical studies, focused on fixed wind turbines, relied on high-order numerical simulations of the Navier-Stokes equations for their superior resolution of the vortex structures. The perturbations were imposed as body forces to the blade tips, and the goal was to investigate the evolution and growth of instabilities. However, since the focus was not the flow around the airfoil blades, the blades were not resolved but modeled with the Actuator Line Method \added{(ALM)~}\citep{sorensen2002numerical}. \added{The ALM has the benefit of not requiring the detailed mesh necessary to fully represent the blade geometry~\citep{sorensen2015simulation}. A more faithful representation of the blades, using immersed boundary methods~\citep{posa2021instability}, observed the same tip vortex instabilities mechanisms. In a follow-up study~\citep{posa2022recovery}, tip vortex instabilities were confirmed as having a major role on the wake recovery of a configuration of two in-line turbines.}

Our recent work \citep{kleine2021stability} connects the study of wakes of heaving turbines and the stability of helical vortices. Using a high-order spectral element method code, we performed numerical simulations of a heaving turbine and qualitatively compared some of the flow features with the predictions of linear stability theory. In the present study, the work of \citet{kleine2021stability} is extended to include other types of motion. Simulations of pitch and surge motions, in addition to heave, are performed. The details of the derivation of the stability analysis for moving turbines are greatly extended to provide more details and to include all the degrees of freedom: surge, heave, sway, pitch, yaw and roll of the turbine. The growth rate and the modes induced by turbine motion from the numerical simulations are quantitatively compared to the predictions of the linear stability theory in the present work. A longer domain is used in this study to analyse the evolution in space of the flow structures created by the turbine motion and its impact on the flow in the region where a downstream turbine would be installed.

The present study focus on the stability of the wake of turbines under motion. Hence, the effect of motion on forces and performance is not discussed. The goal is to provide models based on the theory of hydrodynamic stability that could be applicable to several cases and to validate these models. Therefore, we present the model using non-dimensional and normalized parameters. In such a manner, the models are general and could be applicable to most turbines in different conditions. For completeness, some of the developments, descriptions, discussions and conclusions of \citet{kleine2021stability} are repeated in this work. This article is organized as follows. In section~\ref{sec:methods}, the numerical method used for the simulations is described. A first-order approximation of the stability properties of the flow is presented in section~\ref{sec:stability}. The numerical results are shown and compared to the stability theory in section~\ref{sec:results}. Finally, the main conclusions are presented in section~\ref{sec:conclusions}.

\section{Numerical Methods} \label{sec:methods}

\subsection{Numerical Solver, Domain and Boundary Conditions}

\texttt{Nek5000}, a spectral element method (SEM) code, is used to solve the three-dimensional Navier-Stokes equations in a fixed frame of reference. The spectral element method exhibits low dispersion and dissipation \citep{fischer2015nek5000}, which is relevant for stability calculations. In each spectral element, seventh-order Lagrange polynomials on Gauss-Lobatto-Legendre quadrature points are used for spatial discretization and a third-order implicit/explicit scheme is applied for temporal discretization. To stabilize numerical instabilities, filtering is applied \citep{kleusberg2019wind}.

\begin{figure}[ht]
  \centering
  \begin{tabular}{c c}
    (a) &
    \includegraphics{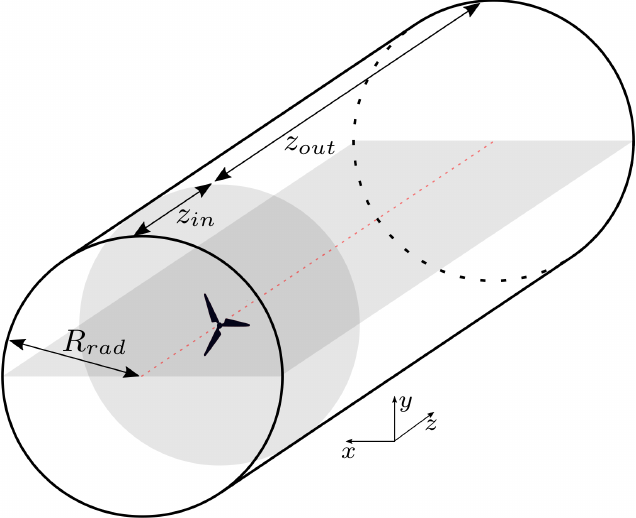} \\
    (b) &
    \includegraphics{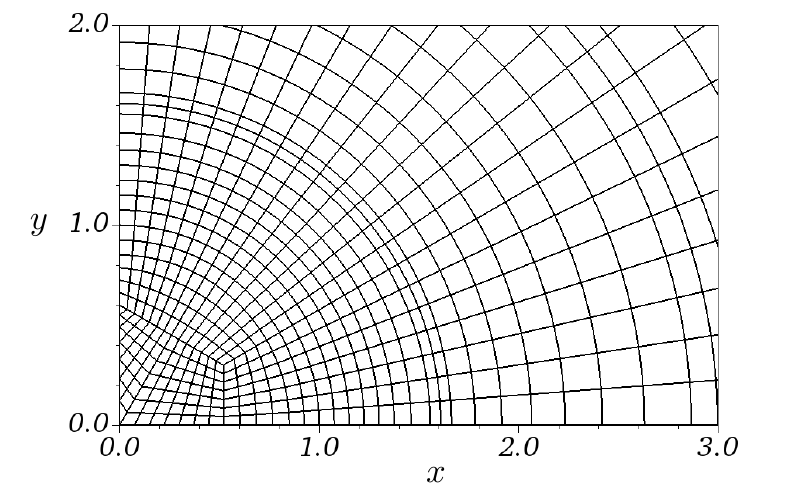} \\
    (c) &
    \includegraphics{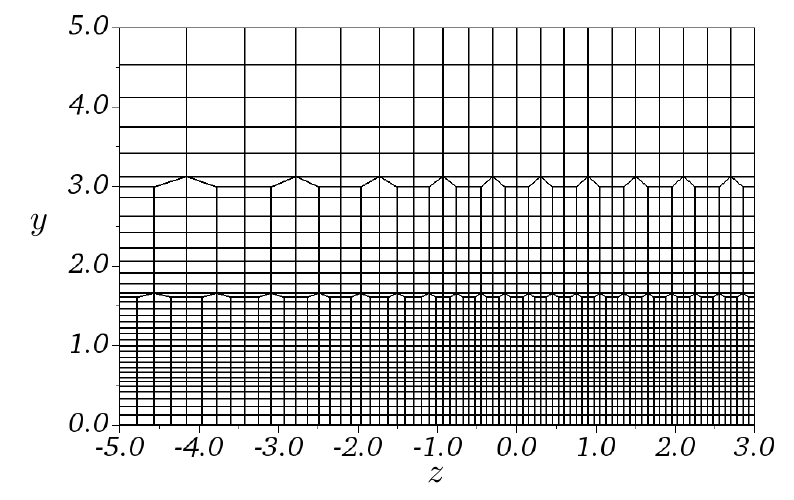}
  \end{tabular}
  \caption{(a) Schematic view of the computational domain. The center of the turbine is located at $(x_{t0},y_{t0},z_{t0})=(0,0,0)$. $R_{rad}=5$ and $z_{in}=5$. All distances are non-dimensionalized by the radius. (b) and (c) Mesh at the center of the domain, showing the spectral elements. Mesh is symmetric with respect to $xz$ and $yz$-planes}
  \label{fig:Schematicdomain}
\end{figure}

All quantities are non-dimensionalized by the turbine radius $R$ and the free-stream velocity $W_\infty$. The turbine is positioned at the origin of a cylindrical domain of radius $R_{rad}=5$ (figure~\ref{fig:Schematicdomain}(a)). The distances of the inlet and the outlet to the turbine are $z_{in}=5$ and $z_{out}=22.10$, respectively, for the longer domain used in the simulations reported in section~\ref{sec:num_results}. Around the center of the domain, the elementwise discretization in the streamwise direction is uniform between $-0.6<z<14.025$, with constant spacing $\Delta z = 0.075$. \added{This discretization is consistent with previous studies\cite{liu2022evaluating}, which evaluated the minimum grid spacing necessary for accurate simulations using actuator line methods, bearing in mind that in this work we are using seventh-order Lagrange polynomials as basis functions.} A shorter domain with $z_{out}=11.78$ and uniform streamwise discretization between $-0.6<z<8.025$ is used for comparison of the growth rate and modes with the stability theory.

As can be seen in figures~\ref{fig:Schematicdomain}(b) and (c), the discretization is coarser in regions farther from the turbine, in order to reduce computational cost. Dirichlet boundary conditions with constant velocity $W_{\infty}=1$ are imposed on the inflow and lateral boundaries. Sheared and turbulent inflow are not considered in this study. At the outlet, the natural outflow boundary condition is imposed in conjunction with a sponge region, with a width $\Delta z=2.5$, that forces the $x$ and $y$ components of velocity to zero. Due to the nature of the outflow boundary condition, no forcing of the $z$-component is needed.

\added{There is a confinement effect for this domain, given the relatively low value of $R_{rad}=5$, which corresponds to a blockage ratio of $(R/R_{rad})^2=4\%$. The results were compared to a simulation with a domain with $R_{rad}=10$ (the domain and grid described in Ref.~\onlinecite{kleine2019tip}), with a blockage ratio of $1\%$. Differences in the order of 1-2\% in the streamwise velocity were observed in regions upstream of the turbine and outside the wake. Inside the wake the differences are smaller. More importantly, the confinement effect did not affect significantly the position of the vortices and the qualitative aspects of the flow. Also, from a theoretical perspective, the influence of the confinement for $R_{rad}=5$ on the stability of the vortices is negligible. Therefore, we concluded that the effect of the confinement on the main results and conclusions is negligible.}

\subsection{Wind Turbine Modelling} \label{sec:turbinemodel}

The wind turbines are modelled using the actuator line method\added{ (ALM)}~\citep{sorensen2002numerical}. In this method, the blades are represented by body forces calculated from airfoil data and local velocity. Considering the local velocity and the chord and twist distributions, the normal and tangential forces are calculated for $N_{\replaced{ACL}{ALM}}$ points ($\mathbf{x}_{n}$) along the blades. The discrete two-dimensional force vector ($\mathbf{F}_{2D}$) is projected on the domain ($\mathbf{x}$) using the convolution of the force with a three-dimensional Gaussian kernel:
\begin{equation}
    \mathbf{F}(\mathbf{x},t)=\mathbf{F}_{2D}(\mathbf{x}_{n},t) \ast \eta_{\epsilon}(|\mathbf{x}-\mathbf{x}_{n}|)
\end{equation}
\begin{equation}
    \eta_{\epsilon}(|\mathbf{x}-\mathbf{x}_{n}|)=\frac{1}{\varepsilon^3 \pi^{3/2} } \exp \left[ - \left( \frac{|\mathbf{x}-\mathbf{x}_{n}|}{\varepsilon} \right)^2 \right] ,
\end{equation}
where $\varepsilon$ is a smearing parameter. Following parametric studies \citep{kleusberg2019wind}, $\varepsilon \approx 3.5 \Delta r$ (where $\Delta r$ is the averaged grid spacing) and each blade is discretized with 100 points.

The turbine is modelled after the turbine of the Blind Test \citep{krogstad2013blind}. The chosen operation condition was the tip-speed ratio $\lambda=\Omega R / W_{\infty}=6$, which corresponds to the optimum performance for this turbine ($\Omega$ is the angular frequency of the rotor). Hub and tower were not included, since the focus of the work is the stability of the tip vortices. The code was extensively validated for this turbine, showing good agreement between numerical simulations and experimental results \citep{kleusberg2017high,muhle2018blind,kleusberg2019wind} and was previously used in vortex stability studies \citep{kleusberg2019tip,kleusberg2019wind,kleine2019tip} with the same turbine and similar discretization and actuator line parameters.

The lift and drag coefficients of the 14\% thick NREL S826 airfoil were obtained by \citet{sarmast2012experimental}. For the computational simulation, a Reynolds number based on the radius of $Re_R=W_{\infty} R/\nu=50,000$ is used (where $\nu$ is the kinematic viscosity). However, since the Reynolds number \replaced{used to interpolate the airfoil data from the corresponding }{of the} lookup table can be independent of the CFD Reynolds number, a constant factor of $5.64$ is applied to the local Reynolds number used to interpolate the airfoil lift and drag coefficients, so that the tip local Reynolds number (based on the chord at the tip, $c_t$) is in the order of $Re_c^{tip}=W_{\infty} \lambda c_t/\nu \approx 10^5$, matching the value from the experiments~\citep{krogstad2013blind}.

\added{The ALM has been shown to model very well the near wake, when compared both to experimental~\citep{nilsson2015validation} and simpler vortex methods~\citep{sarmast2016comparison}. The main observed difference is in the vortex core size~\citep{nilsson2015validation} (see also Refs.~\onlinecite{dag2020new,martinez2019filtered} for an explanation of this difference), which does not have a first-order effect on the stability properties, since the main stability mechanism is inviscid. As can be seen in Refs.~\onlinecite{widnall1972stability,gupta1974theoretical,quaranta2015long}, the effect of the core size on the stability and initial dynamics of the vortices is secondary, through the desingularization of the Biot-Savart law. For this reason, previous studies using the ALM had good agreement with theoretical predictions~\citep{ivanell2010stability,sarmast2014mutual,kleusberg2019tip,kleusberg2019wind} and it is considered an important tool to study the stability of tip vortices~\citep{sorensen2011instability,sorensen2015simulation}.}

\subsection{Turbine Motion} \label{sec:turbinemotion}

\begin{figure}[ht]
  \centering
  \includegraphics{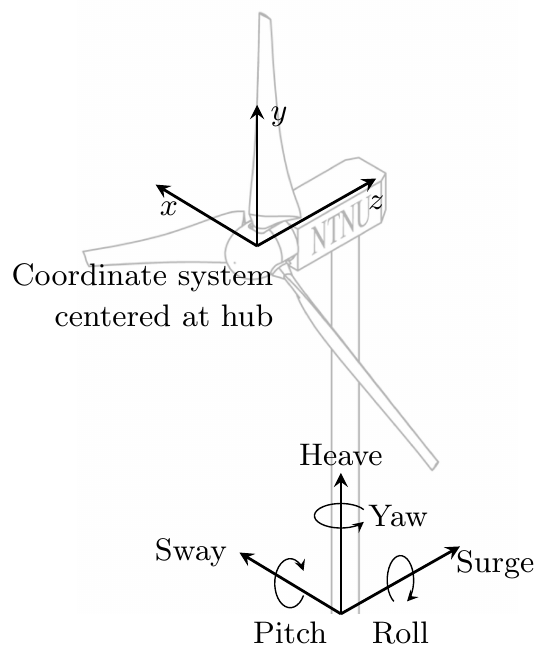}
  \caption{Types of motion of a FOWT. Coordinate system is centered at the hub of the turbine, but the axis of rotation for the turbine motion can be at a different position. Adapted from Ref.~\onlinecite{schottler2017comparative}\added{: J. Schottler, F. Mühle, J. Bartl, J. Peinke, M. S. Adaramola, L. Sætran, and M. Hölling, J. Phys.: Conf. Series 854, 012032 (2017); licensed under a Creative Commons Attribution (CC BY) license}.}
  \label{fig:motionFOWT}
\end{figure}

When a wind turbine is mounted on a floating platform, it is subjected to motions that result from the loads imposed by the wind on the turbine, the waves and sea currents on the platform, and the reaction and constraints of the mooring lines. Wind and waves are usually aligned, and the yaw control system of the turbine guarantees that the rotor faces the wind directly. The floating platforms that are usually employed for floating turbines, namely spar-buoy, semi-submersible platform, and barge, have a high degree of spatial symmetry with respect to the vertical axis. All these aspects make the movement of the turbine occur mostly at the vertical plane parallel to the wind. In other words, the most relevant motions of a floating offshore wind turbine are heave (translation in the $y$-direction), surge (translation in the $z$-direction) and pitch (rotation around the $x$-axis), see figure~\ref{fig:motionFOWT}.

In Ref.~\onlinecite{kleine2021stability} we considered heave motion only. To assess the amplitude and frequency of heave, we performed simulations using the software FAST~v8 \citep{Wendt2016} for the NREL 5MW turbine \citep{jonkman2009definition} mounted on the OC4 semi-submersible platform \citep{Robertson2014}, subject to a \unit[10.3]{m/s} wind speed and waves corresponding to a JONSWAP wave spectrum with significant height (Hs) of \unit[1.5]{m} peak spectral period (Tp) of \unit[8]{s}. These environmental conditions are typical of the Southeast Brazilian continental shelf. Simulations were run considering water depths from \unit[500]{m} to \unit[1000]{m}, with catenary mooring lines designed specifically for each of the depths. We observed heave amplitudes of approximately $1\%$ of the radius ($2\%$ peak-to-peak), and a frequency of about a half of the rotation of the blades. The distance between the center of the turbine and the axis of rotation of pitch was estimated to be between $1.4 R$ and $1.5 R$. So, in the current study, we employ these values as starting points for heave, surge and pitch motions, and investigate how variations in frequency and type of motion affect the wakes produced.

It should be noted that the turbine used to estimate the amplitude and frequency of motion is not the same used in the numerical simulations (described in section~\ref{sec:turbinemodel}). Nevertheless, we assume that the order of magnitude of the non-dimensional parameters of the motion is independent of the turbine.

In the current simulations, the motion and the tip speed ratio of the turbine are prescribed. In other words, the turbine motion does not change the tip speed ratio and the non-steady fluid-dynamic effects do not affect the amplitude and frequency of motion. The heave and surge motions were implemented as the movement of the center $(x_t,y_t,z_t)$ of the actuator lines:
\begin{equation}
    \begin{bmatrix}
      x_t \\
      y_t \\
      z_t
    \end{bmatrix} =
    \begin{bmatrix}
      x_{t0} \\
      y_{t0} + A_y \sin{(\omega_y t)} \\
      z_{t0} + A_z \sin{(\omega_z t)}
    \end{bmatrix}
\end{equation}
where $\omega$ is the angular frequency of motion and the subscripts $_y$ and $_z$ represent heave and surge, respectively. We define $\omega^*=\omega/\Omega$ as the normalized angular frequency.

Pitch motion, defined as rotation along the $x$-axis, is implemented by changing both the position of the center of the turbine and the angle of the turbine with respect to the freestream. The distance of the center of the turbine to the axis of pitch rotation in the $z$-direction is considered to be negligible compared to the distance in the $y$-direction and to the radius of the turbine. The center of pitch rotation is assumed to be at position $R_p$, at the $y$-axis. The equation that defines the position of the tip of the blades $(x_b,y_b,z_b)$ under pitch motion (subscript $_p$) is
\begin{widetext}
\begin{equation}
    \begin{bmatrix}
      x_b \\
      y_b \\
      z_b
    \end{bmatrix} =
    \begin{bmatrix}
      x_{t0} + R \cos(\Omega t+\phi_n) \\
      y_{t0} + R \sin(\Omega t+\phi_n) - (z_{t0}-R_p + R \sin(\Omega t+\phi_n))(1-\cos(A_p \sin{(\omega_p t)})) \\
      z_{t0} + (z_{t0}-R_p + R \sin(\Omega t+\phi_n)) \sin(A_p \sin{(\omega_p t)})
    \end{bmatrix}
    \label{eq:bladepitch}
\end{equation}
\end{widetext}
where $\phi_n=2\pi n /N_b$ is the initial azimuthal position of each blade $n$ ($n=0$ to $N_b-1$, where $N_b=3$ is the number of blades) and $A_p$ is the angular amplitude of motion, measured in radians. According to the estimates performed, the position of the axis of rotation of pitch is rounded to $R_p=-1.5R$.

Sway (translation in the $x$-direction), yaw (rotation around the $y$-axis) and roll (rotation around the $z$-axis) are not simulated. However, the first-order approximation of the perturbations and discussion on the stability of these motions are presented in section~\ref{sec:stability} and appendix \ref{app:perturbationyr}.

\section{Stability analysis} \label{sec:stability}
In this section, a linear stability analysis of the flow is presented, based on a first-order approximation of the perturbations imposed by the motion of the turbine. This was first presented for the heave motion in Ref.~\onlinecite{kleine2021stability} and is extended here for the other degrees of freedom. Initially, first-order approximations of the perturbations on the helical vortex system caused by the turbine motion are obtained for all types of motion. The perturbations are then compared to the unstable modes predicted by the classical results of the linear stability of a two-dimensional infinite row of vortices \citep{lamb1932hydrodynamics} and perfect helical vortices \citep{widnall1972stability,gupta1974theoretical}.

\subsection{Perturbations imposed by turbine motion}

\subsubsection{Reference system for the stability analysis} \label{sec:perturbationref}

For a fixed-bottom turbine, the equation that defines the tip of the blade that rotates in the positive azimuthal direction is
\begin{equation}
    \begin{bmatrix}
      x_{b0} \\
      y_{b0} \\
      z_{b0}
    \end{bmatrix} =
    \begin{bmatrix}
      R \cos(\Omega t+\phi_n) \\
      R \sin(\Omega t+\phi_n) \\
      0
    \end{bmatrix}
    \label{eq:bladepos0}
\end{equation}
where $(x_{t0},y_{t0},z_{t0})=(0,0,0)$ is assumed without loss of generality.

Previous studies identified that the tip vortices are mainly advected by the flow in the streamwise direction \citep{kleusberg2019tip}, hence, a uniform advection velocity $w_c$ in the streamwise direction is considered. The position $(x_{v0}',y_{v0}',z_{v0}')$ of the tip vortices emitted by the blades of equation (\ref{eq:bladepos0}) is
\begin{equation}
    \begin{bmatrix}
      x_{v0}' \\
      y_{v0}' \\
      z_{v0'}
    \end{bmatrix} =
    \begin{bmatrix}
      R \cos(\Omega \tau_0+\phi_n) \\
      R \sin(\Omega \tau_0+\phi_n) \\
      w_c (t-\tau_0)
    \end{bmatrix}
    \label{eq:vortpos0ref0}
\end{equation}
where $\tau_0$ is the time at which the vortex particle was emitted. 

Imposing sway, heave and surge (subscripts $_x$, $_y$ and $_z$, respectively) to the turbine, the position of the tip of the blades is
\begin{equation}
    \begin{bmatrix}
      x_{b} \\
      y_{b} \\
      z_{b}
    \end{bmatrix} =
    \begin{bmatrix}
      R \cos(\Omega t+\phi_n) + A_x \sin{(\omega_x \tau)} \\
      R \sin(\Omega t+\phi_n) + A_y \sin{(\omega_y \tau)} \\
      A_z \sin{(\omega_z t)}
    \end{bmatrix} .
    \label{eq:bladepos}
\end{equation}

The turbine motion imposes perturbations to the position of the vortices, that grow in time and space. In order to study the evolution of these perturbations, we initially do not consider this growth, assuming only advection of the vortices. The position $(x_{v}',y_{v}',z_{v}')$ of the tip vortices with turbine motion is then
\begin{equation}
    \begin{bmatrix}
      x_v' \\
      y_v' \\
      z_v'
    \end{bmatrix} =
    \begin{bmatrix}
      R \cos(\Omega \tau_0+\phi_n) + A_x \sin{(\omega_x \tau_0)} \\
      R \sin(\Omega \tau_0+\phi_n) + A_y \sin{(\omega_y \tau_0)} \\
      w_c (t-\tau_0) + A_z \sin{(\omega_z \tau_0)}
    \end{bmatrix} .
    \label{eq:vortposref0}
\end{equation}

In the general case, there would be a phase difference between the motions and they could be composed by several frequencies. However, as a linear approximation, these motions do not interact, so the phase difference is not relevant for the present analysis. Therefore, in favor of a more concise notation, the phase is omitted and only one frequency for each motion is analysed.

The helical structure of the vortices of equation~(\ref{eq:vortpos0ref0}) change in time. The vortical particles translate in $z$, making the helical structure translate or rotate, depending on the point of view (for further discussion about this topic, see~\citet{okulov2020self}). In order to compare with classical stability results \citep{lamb1932hydrodynamics,widnall1972stability,gupta1974theoretical}, a frame of reference in which the vortices are static should be chosen. There are infinite choices for the pair of streamwise and azimuthal velocities that make the helical structure static, but different reference systems give different frequencies in the stability analysis \citep{brynjell2020numerical}. We then choose the reference system that follows the vortical particles, which is the same as that employed by \citet{widnall1972stability,gupta1974theoretical} and \citet{lamb1932hydrodynamics}. Hence, considering the assumption that the vortices are simply advected by a constant streamwise velocity, the reference system is a translating frame of reference with velocity $w_c$. Taking the origin in $z$ of the system as the position where $\tau_0=0$, equation (\ref{eq:vortposref0}) becomes
\begin{equation}
    \begin{bmatrix}
      x_v'' \\
      y_v'' \\
      z_v''
    \end{bmatrix} =
    \begin{bmatrix}
      R \cos(\Omega \tau_0+\phi_n) + A_x \sin{(\omega_x \tau_0)} \\
      R \sin(\Omega \tau_0+\phi_n) + A_y \sin{(\omega_y \tau_0)} \\
      - w_c \tau_0 + A_z \sin{(\omega_z \tau_0)}
    \end{bmatrix} .
    \label{eq:vortposref1}
\end{equation}
Hence $\tau_0$ can be thought of as the parameter that defines the helical vortices.

It is worth noting that the helical vortices defined by equation (\ref{eq:vortposref1}) are left-handed helices, just as the vortices generated in our simulations. However, most of the stability studies consider right-handed helices, including our implementation of the method of \citet{gupta1974theoretical}. Also, for positive $w_c$, the values of $z_v$ and $\tau_0$ have different signs, which indicates that the perturbation would have a different sign when written in function of $z$: $A \sin{(\omega \tau_0)} = - A \sin{(\omega z_{v0}/w_c)}$. Both of these characteristics are undesirable for the comparison of the perturbations with the stability theory. Hence, without loss of generality, we consider a modification to the vortex system, so the position is given by
\begin{equation}
    \begin{bmatrix}
      x_v \\
      y_v \\
      z_v
    \end{bmatrix} =
    \begin{bmatrix}
      R \cos(\Omega \tau+\phi_n) + A_x \sin{(\omega_x \tau)} \\
      R \sin(\Omega \tau+\phi_n) + A_y \sin{(\omega_y \tau)} \\
      w_c \tau + A_z \sin{(\omega_z \tau)}
    \end{bmatrix} 
    \label{eq:vortposref}
\end{equation}
which has the desirable properties of being formed by right-handed helices and having perturbations that have a sinusoidal component in $z$ with the same sign of the new parameter $\tau$, providing greater similarity to the geometry and perturbations of classical stability studies. The stability properties for the geometry of equation~(\ref{eq:vortposref}) are the same of equation~(\ref{eq:vortposref1}), but the comparison with classical results can be performed without further modification to the reference system by using equation~(\ref{eq:vortposref}). On the other hand, comparison to the simulations should be made considering the different reference systems.

\subsubsection{First-order approximation of perturbation imposed by sway, heave and surge} \label{sec:perturbationshs}

Using the reference system of equation~(\ref{eq:vortposref}), the position of the tip vortices of right-handed helices without any turbine motion would be
\begin{equation}
    \begin{bmatrix}
      x_{v0} \\
      y_{v0} \\
      z_{v0}
    \end{bmatrix} =
    \begin{bmatrix}
      R \cos(\Omega \tau+\phi_n) \\
      R \sin(\Omega \tau+\phi_n) \\
      w_c \tau
    \end{bmatrix} .
    \label{eq:vortpos0}
\end{equation}
Thus, the perturbation directly induced by the turbine motion is \comment{Added break}$(\delta x_v,\delta y_v,\delta z_v) = (A_x \sin{(\omega_x \tau)}, \allowbreak A_y \sin{(\omega_y \tau)}, A_z \sin{(\omega_z \tau)})$, in cartesian coordinates. For $A_x,A_y \ll R$, first-order approximations of the perturbation in the radial direction and azimuthal directions are 
\begin{widetext}
\begin{equation}
  \begin{aligned}
    \delta r_v & = \sqrt{x_v^2+y_v^2}-R \\ & \approx \sqrt{R^2+2 R A_y \sin(\Omega \tau+\phi_n) \sin{(\omega_y \tau)}+2 R A_x \cos(\Omega \tau+\phi_n) \sin{(\omega_x \tau)}}-R \\ 
    & \approx A_y \sin(\Omega \tau+\phi_n) \sin{(\omega_y \tau)} + A_x \cos(\Omega \tau+\phi_n) \sin{(\omega_x \tau)}, 
  \end{aligned}
\end{equation}
\begin{equation}
    \delta \varphi_v = \arctan{\frac{x_v}{y_v}} - (\Omega \tau+\phi_n) \approx - \frac{A_y}{R} \cos(\Omega \tau+\phi_n) \sin{(\omega_y \tau)} + \frac{A_x}{R} \sin(\Omega \tau+\phi_n) \sin{(\omega_x \tau)} .
\end{equation}
\end{widetext}

The order of magnitude of $\delta r_v$ and $R \delta \varphi_v$ is the same. However, the perturbation in the azimuthal direction is not usually as relevant as the other components for the stability of helical vortices of low pitch ($h/R \ll 1 \iff$ $w_c/(N_b \Omega R) \ll 1$), so, as an approximation, it is neglected. Defining $\theta = \Omega \tau+\phi_n = \Omega z_{v0}/w_c+\phi_n$, the first-order approximation of the perturbation in the radial and streamwise directions are defined for the helices in function of the azimuthal angle as
\begin{equation}
  \begin{aligned}
    &
    \begin{bmatrix}
      \delta r_v  \\
      \delta z_v 
    \end{bmatrix} = \\ &
    \begin{bmatrix}
      A_y \sin{(\omega_y^*( \theta-\phi_n))} \sin \theta + A_x \sin{(\omega_x^*( \theta-\phi_n))} \cos \theta  \\
      A_z \sin{(\omega_z^*( \theta-\phi_n))}
    \end{bmatrix}
  \end{aligned}
    \label{eq:perturbations}
\end{equation}

This initial perturbation induced by the motion grows in time and in space, as the vortices move downstream. The growth rate can be estimated using analytical stability models. The applied stability models consider infinite uniform vortices, which is a limitation of the model developed here. Hence, the region of wake expansion is not considered but wake expansion in the region of uniform helices could be partially taken into account by using a different value of $R$.

Another limitation is that only perturbations in position are considered. Other effects may be also relevant for some types of motion. For example, for heave and sway motion, the change in position of blockage effect of the turbine might impose other perturbations to the wake that are not considered here. The modification of the forces at the blades and, consequently, the circulation is also not accounted. If these effects also have the form of equation~(\ref{eq:perturbations}) (pure sinusoidal oscillation for surge and sinusoidal with $\sin \theta$ and $\cos \theta$ envelopes for heave and sway), then the analysis performed here would also be applicable.

\subsubsection{First-order approximation of perturbation imposed by pitch} \label{sec:perturbationpwr}

From equation~(\ref{eq:bladepitch}), applying the change in reference system discussed in section~\ref{sec:perturbationref}, the equation that defines the position of the tip vortices under pitch motion becomes
\begin{widetext}
\begin{equation}
  \begin{aligned}
    \begin{bmatrix}
      x_v \\
      y_v \\
      z_v
    \end{bmatrix} & =
    \begin{bmatrix}
      R \cos(\Omega \tau+\phi_n) \\
      R \sin(\Omega \tau+\phi_n) - (-R_p + R \sin(\Omega \tau+\phi_n))(1-\cos(A_p \sin{(\omega_p \tau)})) \\
      w_c \tau + (-R_p + R \sin(\Omega \tau+\phi_n)) \sin(A_p \sin{(\omega_p \tau)})
    \end{bmatrix} \\
    & \approx     \begin{bmatrix}
      R \cos(\Omega \tau+\phi_n) \\
      R \sin(\Omega \tau+\phi_n)+(-R_p + R \sin(\Omega \tau+\phi_n))\frac{(A_p \sin{(\omega_p \tau)})^2}{2} \\
      w_c \tau + (-R_p + R \sin(\Omega \tau+\phi_n)) A_p \sin{(\omega_p \tau)}
    \end{bmatrix}
  \end{aligned}
\end{equation}
\end{widetext}
where the approximation comes by assuming that the maximum pitch angle, $A_p$, is low.
We can directly notice that the motion in the $y$-direction is of second order. For low amplitudes, this can be neglected and the first-order approximation of the perturbation due to pitch motion becomes
\begin{equation}
  \begin{aligned} &
    \begin{bmatrix}
      \delta x_v \\
      \delta y_v \\
      \delta z_v
    \end{bmatrix} =
    \begin{bmatrix}
      0 \\
      0 \\
      -R_p A_p \sin{(\omega_p \tau)} + R A_p \sin(\Omega \tau+\phi_n) \sin{(\omega_p \tau)}
    \end{bmatrix} \\ & =
    \begin{bmatrix}
      0 \\
      0 \\
      -R_p A_p \sin{(\omega_p^*( \theta-\phi_n))} + R A_p \sin{(\omega_p^*( \theta-\phi_n))} \sin \theta 
    \end{bmatrix}
  \end{aligned}
\end{equation}
which, written in the radial and streamwise directions, is
\begin{equation}
  \begin{aligned} &
    \begin{bmatrix} 
      \delta r_v  \\
      \delta z_v 
    \end{bmatrix} = \\ &
    \begin{bmatrix}
      0 \\
      -R_p A_p \sin{(\omega_p^*( \theta-\phi_n))} + R A_p \sin{(\omega_p^*( \theta-\phi_n))} \sin \theta 
    \end{bmatrix} .
  \end{aligned}
    \label{eq:perturbationspithc}
\end{equation}
Hence, the perturbations imposed by the pitch motion are equivalent to perturbations imposed by the surge motion. For $R_p \gg R$ this corresponds to a surge motion with amplitude $R_p A_p$. For $R_p \ll R$ this corresponds to a motion in $z$-direction composed of component $\sin{(\omega_p^*( \theta-\phi_n))} \sin \theta$ with amplitude $R A_p$, which has the same sinusoidal envelope of the heave component in equation~(\ref{eq:perturbations}).

The first-order approximations of the perturbations imposed by yaw and roll are shown in appendix~\ref{app:perturbationyr}. The change in the direction of the thrust vector in the case of yaw or pitch, that causes wake deflection (see section 4.2 of \citet{burton2011wind}), could potentially also induce perturbations to the vortices, including in the radial direction. These effects are not considered in this small-perturbation model.

As can be observed, all the components of the perturbations induced by pitch, yaw and roll have the same format of an amplitude multiplied by $(\sin{(\omega^*( \theta-\phi_n))} \sin \theta)$, $(\sin{(\omega^*( \theta-\phi_n))} \cos \theta)$ or $(\sin{(\omega^*( \theta-\phi_n))})$. Conceptually, for the purposes of the analysis of the following sections, these components are equivalent to the perturbations of equation~(\ref{eq:perturbations}). Hence, the detailed analyses are restricted to the case of surge and heave. In the end, a parallel to the case of other kinds of motion is made.

\subsection{Stability of a 2-d Row of Vortices} \label{sec:stability2d}

\begin{figure*}[t]
  \centering
  \includegraphics{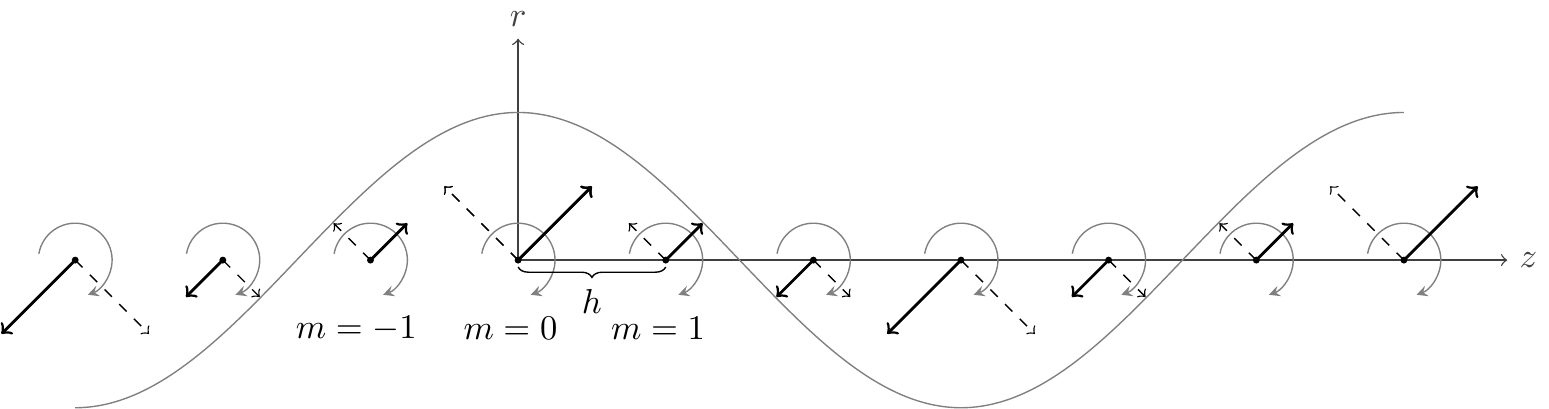}
  \caption{Real part of the modes predicted by the linear stability theory for a row of vortices, for $p=1/6$ ($\omega^*=p N_b=0.5$). Stable mode is indicated by dashed arrows and unstable mode is indicated by solid arrows.}
  \label{fig:vortexrow}
\end{figure*}

Several studies made the connection between the stability of helical vortices and a two-dimensional row of vortices~\citep{leweke2014long,sarmast2014mutual,quaranta2015long,quaranta2019local}, showing that the scaled growth rate of \added{long-wave and mutual inductance} instabilities collapses to the theoretical value for an infinite row of vortices calculated by \citet{lamb1932hydrodynamics} (see also Refs.~\onlinecite{saffman1993vortex,kleine2022stability}). The growth rate $\sigma$ is given by \citep{saffman1993vortex,kleine2022stability}:
\begin{equation}
    \sigma = \pm \frac{\pi \Gamma}{h^2} p (1-p)
  \label{eq:growth2d}
\end{equation}
where $h$ is the distance between neighboring vortices, $\Gamma$ is the circulation and $p$ is a dimensionless subharmonic wavenumber (see figure~\ref{fig:vortexrow}). The maximum growth rate occurs for $p=1/2$, which corresponds to the vortex pairing where neighboring vortices are out-of-phase.

The growth rate when scaled using the distance of neighboring vortices and the circulation, $\tilde{\sigma}=\sigma (2h^2/\Gamma)$, has a maximum value of $\pi/2$. The growth rate of the vortex pairing mode collapse to $\pi/2$ even for non-uniform helices created by turbines under sheared inflow or yawed, if scaled by local properties, as shown by \citet{kleusberg2019tip} and \citet{kleusberg2019wind}. The two modes $(\delta r_{m_{2d}1}, \delta z_{m_{2d}1})$ and $(\delta r_{m_{2d}2}, \delta z_{m_{2d}2})$ predicted by the linear stability theory for each $p$ are~\citep{saffman1993vortex,kleine2022stability}
\begin{equation}
    \begin{bmatrix}
      \delta r_{m_{2d}1}  \\
      \delta z_{m_{2d}1} 
    \end{bmatrix} =
    \begin{bmatrix}
      1
      \\
      1
    \end{bmatrix} e^{i 2 \pi p m}
    \text{\quad and \quad}
    \begin{bmatrix}
      \delta r_{m_{2d}2}  \\
      \delta z_{m_{2d}2} 
    \end{bmatrix} =
    \begin{bmatrix}
      1
      \\
      -1
    \end{bmatrix} e^{i 2 \pi p m},
  \label{eq:modes2d}
\end{equation}
where $i$ is the complex unit. These modes are illustrated in figure~\ref{fig:vortexrow}. One of these modes is exponentially damped and the other grows exponentially, depending on the sign of the circulation and the reference system. For the reference system of figure~\ref{fig:vortexrow}, which corresponds to the reference system used in this work for a wind turbine, mode $(\delta r_{m_{2d}1}, \delta z_{m_{2d}1})$ is unstable.

In order to compare to a row of two-dimensional vortices, we take a cross-section of the helical vortices, taking the values of $\theta=\theta_0+2 \pi j$, where $j$ is an integer and $\theta_0$ is chosen to define a cross-section that contains the row of vortices ($0 \leq \theta_0 < 2 \pi$). This is a slightly different approach from that in \citet{quaranta2015long}, where the minimum distance between neighboring vortices was used. However, for helices of low pitch, the difference between the minimum distance and the distance defined by a cross-section is of second order. From equation~(\ref{eq:perturbations}), the displacement of the vortices due to surge, heave and sway along the cross-section defined by $\theta_0$ is
\begin{widetext}
\begin{equation}
  \begin{aligned}
    \begin{bmatrix}
      \delta r_{v_{2d}}  \\
      \delta z_{v_{2d}} 
    \end{bmatrix} & =
    \begin{bmatrix}
      A_y \sin{(\omega_y^* \theta_0+\omega_y^*(2 \pi j-\phi_n))} \sin\theta_0 + A_x \sin{(\omega_x^* \theta_0+\omega_x^*(2 \pi j-\phi_n))} \cos\theta_0  \\
      A_z \sin{(\omega_z^* \theta_0+\omega_z^*(2 \pi j-\phi_n))}
    \end{bmatrix}
    \\
    & =
    \begin{bmatrix}
      (A_y \sin\theta_0) \sin{\left(\omega_y^* \theta_0+\frac{\omega_y^*}{N_b}2 \pi (N_b j-n)\right)} + (A_x \cos\theta_0) \sin{\left(\omega_x^* \theta_0+\frac{\omega_x^*}{N_b}2 \pi (N_b j-n)\right)}
      \\
      A_z \sin{\left(\omega_z^* \theta_0+\frac{\omega_z^*}{N_b}2 \pi (N_b j-n)\right)}
    \end{bmatrix}
    \\
    & =
    \begin{bmatrix}
      (A_y \sin\theta_0) \sin{\left(\omega_y^* \theta_0+\frac{\omega_y^*}{N_b}2 \pi m\right)} + (A_x \cos\theta_0) \sin{\left(\omega_x^* \theta_0+\frac{\omega_x^*}{N_b}2 \pi m\right)} 
      \\
      A_z \sin{\left(\omega_z^* \theta_0+\frac{\omega_z^*}{N_b}2 \pi m\right)}
    \end{bmatrix},
  \end{aligned}
  \label{eq:perturbationsrow}
\end{equation}
\end{widetext}
where the term $N_b j-n$, \replaced{that}{which} can be any integer, has been reinterpreted as the index $m$ of each 2-d vortex (figure \ref{fig:vortexrow}).

Projecting the terms of equation~(\ref{eq:perturbationsrow}) onto the modes defined in equation~(\ref{eq:modes2d}), the perturbation $(\delta r_{v_{2d}}, \delta z_{v_{2d}})$ can be decomposed in the components $(\delta r_{v_{2d}1}, \delta z_{v_{2d}1})$ and $(\delta r_{v_{2d}2}, \delta z_{v_{2d}2})$:
\begin{widetext}
\begin{equation}
  \begin{aligned}
    \begin{bmatrix}
      \delta r_{v_{2d}1}  \\
      \delta z_{v_{2d}1} 
    \end{bmatrix} & =
    \begin{bmatrix}
      1
      \\
      1
    \end{bmatrix} \frac{1}{2} \biggl( (A_y \sin\theta_0) \sin{\left(\omega_y^* \theta_0+\frac{\omega_y^*}{N_b}2 \pi m\right)} + (A_x \cos\theta_0) \sin{\left(\omega_x^* \theta_0+\frac{\omega_x^*}{N_b}2 \pi m\right)} \\ & + A_z \sin{\left(\omega_z^* \theta_0+\frac{\omega_z^*}{N_b}2 \pi m\right)} \biggr) \\
    & =
    \begin{bmatrix}
      1
      \\
      1
    \end{bmatrix} \frac{-1}{4} \left( (A_y \sin\theta_0) i e^{i \omega_y^* \theta_0} e^{i \frac{\omega_y^*}{N_b}2 \pi m} + (A_x \cos\theta_0) i e^{i \omega_x^* \theta_0} e^{i \frac{\omega_x^*}{N_b}2 \pi m} + A_z i e^{i \omega_z^* \theta_0} e^{i \frac{\omega_z^*}{N_b}2 \pi m} \right) + c.c. \\
  \end{aligned}
  \label{eq:perturbations2dproject}
\end{equation}
and
\begin{equation}
  \begin{aligned}
    \begin{bmatrix}
      \delta r_{v_{2d}2}  \\
      \delta z_{v_{2d}2} 
    \end{bmatrix} & =
    \begin{bmatrix}
      1
      \\
      -1
    \end{bmatrix} \frac{1}{2} \biggl( (A_y \sin\theta_0) \sin{\left(\omega_y^* \theta_0+\frac{\omega_y^*}{N_b}2 \pi m\right)} + (A_x \cos\theta_0) \sin{\left(\omega_x^* \theta_0+\frac{\omega_x^*}{N_b}2 \pi m\right)} \\ & - A_z \sin{\left(\omega_z^* \theta_0+\frac{\omega_z^*}{N_b}2 \pi m\right)} \biggr) \\
    & =
    \begin{bmatrix}
      1
      \\
      -1
    \end{bmatrix} \frac{-1}{4} \left( (A_y \sin\theta_0) i e^{i \omega_y^* \theta_0} e^{i \frac{\omega_y^*}{N_b}2 \pi m} + (A_x \cos\theta_0) i e^{i \omega_x^* \theta_0} e^{i \frac{\omega_x^*}{N_b}2 \pi m} - A_z i e^{i \omega_z^* \theta_0} e^{i \frac{\omega_z^*}{N_b}2 \pi m} \right) + c.c.
  \end{aligned}
  \label{eq:perturbations2dproject2}
\end{equation}
\end{widetext}
where the notation $c.c.$ indicates the complex conjugate of the previous term. For the current reference system, mode $1$ grows exponentially and mode $2$ decays. Hence, after some distance from the turbine, mode $(\delta r_{v_{2d}1}, \delta z_{v_{2d}1})$ tends to dominate. It can be noted that there is a symmetry between the perturbation in the component $\delta r_{v_{2d}}$ and $\delta z_{v_{2d}}$: a perturbation in $\delta r_{v_{2d}}$ induces a component in $\delta z_{v_{2d}}$ and vice-versa, because the eigenvectors have equal components in $r$ and $z$.

Qualitatively, the three terms of equation~(\ref{eq:perturbations2dproject}) are similar. The terms $(A_y \sin\theta_0)$, $(A_x \cos\theta_0)$ and $A_z$ are the amplitudes of the perturbation induced by each motion. The main difference is that the amplitude for the heave and sway motion is modulated by the angle $\theta_0$ while the amplitude of the surge motion is uniform in the azimuthal direction. This implies that the heave motion has a maximum amplitude at angles $\theta_0=\pm \pi/2$ and does not induce instability at angles $\theta_0=0$ and $\pi$. Due to the three-dimensionality of the problem and non-linear dynamics, the helical vortices created by a heaving turbine would also interact at angles $\theta_0=0$ and $\pi$. Nevertheless, the vortices at $\pm \pi/2$ would have noticeable larger amplitudes.

\begin{figure}[ht]
    \centering
    \includegraphics[width=0.45\textwidth]{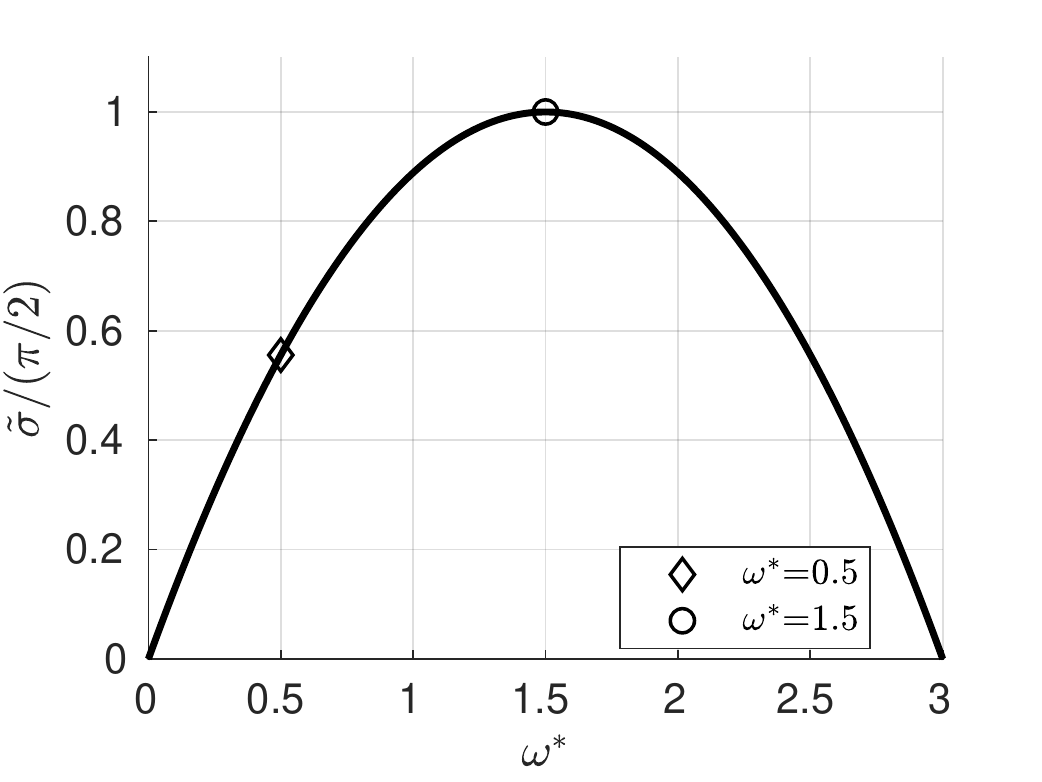}
    \caption{Theoretical growth rate of unstable modes of a two-dimensional row of vortices. In order to make it comparable to helical vortices, the $x$-axis is scaled using the relationship $\omega^*=p N_b$. For $\omega^*>3$ or negative, we consider the value of $\omega^*-j N_b$ where $j$ is any integer. Symbols correspond to modes relevant to cases $\omega^*=0.5$ and $\omega^*=1.5$.}
    \label{fig:eigvalsrow}
\end{figure}

From the terms of equation~(\ref{eq:perturbations2dproject}), it follows directly that the terms $i\omega^* \theta_0$ correspond to a phase shift, and, most importantly, $\omega^*/N_b$ is the dimensionless subharmonic wavenumber $p$ for each motion type. Applying the scaling $p = \omega^*/N_b$ to equation~(\ref{eq:growth2d}), we obtain the theoretical growth rate for a row of vortices shown in figure~\ref{fig:eigvalsrow}. Equation~(\ref{eq:growth2d}), in principle, is only valid for $0 \le p \le 1$. For values of $\omega^*/N_b$ that are outside this interval, however, we use $p = \omega^*/N_b + j$ where $j$ is any integer that makes $0 \le p \le 1$ (including for negative values of $\omega^*$).

As a linear approximation, this growth rate is applicable to all motion types, including pitch, yaw and roll, following the discussion of section~\ref{sec:perturbationpwr}. Because this is a linear analysis both in the approximation of the perturbation and in the stability theory, the perturbations grow independently of each other. The effect of non-linear interaction of the perturbations imposed by different motions is not considered in this theory.

\subsection{Stability of Helical Vortices}

Regarding the stability of the three-dimensional helical vortices, the knowledge of the wavenumbers induced by the perturbations is essential to compare to classical results from the analytical linear stability \citep{widnall1972stability,gupta1974theoretical}. As discussed in \citet{kleusberg2019tip} and \citet{kleine2019tip}, for perturbations imposed as body forces at the tip of the blades, the angular frequency is related to the wavenumber according to the approximate relation $\omega=k\Omega$. However, as it is going to be shown, in the case of perturbations imposed by turbine motion, other wavenumbers can be induced for a certain frequency, depending on the motion of the turbine. 

In the two-dimensional case, the perturbation in one direction, when projected onto the eigenvectors, creates a perturbation in the other component with equal magnitude (section~\ref{sec:stability2d}). This means that the heave and sway motions induce a component in the streamwise direction and the surge motion induces a component in the radial direction, so that $\delta z = \delta r$. In the case of three-dimensional vortices, the most amplified modes predicted by the stability theory also have $\delta z \approx \delta r$, so that the projection of the perturbations onto the eigenvectors would create perturbations in the other component. Hence, we can assume that the perturbation given by equation~(\ref{eq:perturbations}) is composed by two terms
\begin{eqnarray}
    \begin{bmatrix}
      \delta r_{v1}  \\
      \delta z_{v1} 
    \end{bmatrix} =
    \begin{bmatrix}
      1
      \\
      1
    \end{bmatrix} \frac{1}{2} \biggl( & \quad A_y \sin{(\omega_y^*( \theta-\phi_n))} \sin \theta \nonumber\\ & + A_x \sin{(\omega_x^*( \theta-\phi_n))} \cos \theta \nonumber\\ & + A_z \sin{(\omega_z^*( \theta-\phi_n))} \biggr)
    \label{eq:distz} 
\end{eqnarray}
and 
\begin{eqnarray}
    \begin{bmatrix}
      \delta r_{v2}  \\
      \delta z_{v2} 
    \end{bmatrix} =
    \begin{bmatrix}
      1
      \\
      - 1
    \end{bmatrix} \frac{1}{2} \biggl( & \quad A_y \sin{(\omega_y^*( \theta-\phi_n))} \sin \theta \nonumber\\ & + A_x \sin{(\omega_x^*( \theta-\phi_n))} \cos \theta \nonumber\\ &  - A_z \sin{(\omega_z^*( \theta-\phi_n))} \biggr),
    \label{eq:distz2} 
\end{eqnarray}
where $(\delta r_{v1},\delta z_{v1})$ grows and $(\delta r_{v2},\delta z_{v2})$ decays in time and space (for the current choice of reference system).

Using trigonometric identities, equation~(\ref{eq:distz}) can be decomposed in
\begin{eqnarray}
%  \begin{aligned}
    \begin{bmatrix}
      \delta r_{v1}  \\
      \delta z_{v1} 
    \end{bmatrix} =
    \begin{bmatrix}
      1
      \\
      1
    \end{bmatrix} \frac{1}{2} \biggl( & \quad \frac{A_y}{2} \bigl( & \cos{((\omega_y^*-1) (\theta-\phi_n) - \phi_n)} \nonumber\\ & &- \cos{((\omega_y^*+1) (\theta-\phi_n) + \phi_n)} \bigr) \biggr. \nonumber\\
    & + \frac{A_x}{2} \bigl(&\sin{((\omega_x^*-1) (\theta-\phi_n) - \phi_n)} \nonumber\\ & &+  \sin{((\omega_x^*+1) (\theta-\phi_n) + \phi_n)} \bigr) \nonumber\\
    & \biggl. + A_z & \sin{(\omega_z^* (\theta-\phi_n))} \biggr) 
%  \end{aligned}
  \label{eq:distz_comp}
\end{eqnarray}
and analogously for equation~(\ref{eq:distz2}). The wavenumber $k$ of each perturbation is given by the term that multiplies $\theta$, while the terms $\pm\phi_n$ are phase shifts between the different vortices $n$. Thus, a sway or heave motion ($x$ and $y$-components) with frequency $\omega^*$ creates perturbations with wavenumbers $k=\omega^*-1$ and $k=\omega^*+1$. The surge motion ($z$-component), on the other hand, creates perturbations with wavenumber $k=\omega^*$. The surge motion is similar to previous studies where body forces were used as perturbations at the tip, which arrived at the same approximated relation $k=\omega^*$ \citep{kleusberg2019tip}.

\begin{figure*}[ht]
    \centering
    \includegraphics[width=0.7\textwidth]{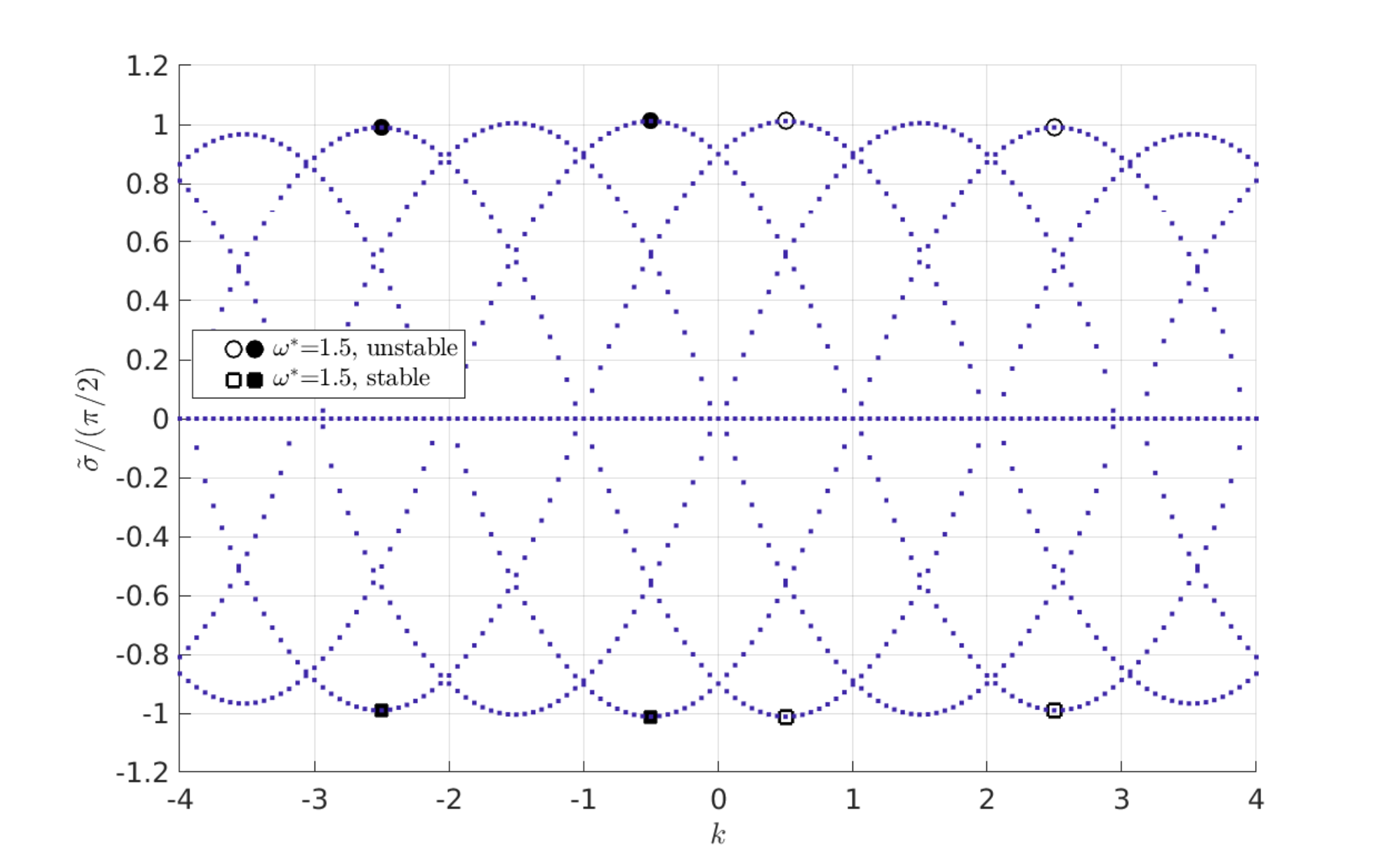}
    \caption{Theoretical growth rate for the eigenvalues as function of the wavenumber for a triple helix, obtained using the method of \citet{gupta1974theoretical}. Symbols correspond to modes relevant to heave case with $\omega^*=1.5$. Primary modes are shown using open symbols and their complex conjugate are shown with filled symbols.}
    \label{fig:eigvalsfull}
\end{figure*}

Analogously to the row of vortices, when the real perturbation is projected onto the eigenvectors, a pair of complex conjugate modes is excited, see equations~(\ref{eq:perturbations2dproject}) and (\ref{eq:perturbations2dproject2}). Therefore, the heave and sway motions excite modes with wavenumbers $k=\pm(\omega^*-1)$ and $k=\pm(\omega^*+1)$ and the surge motion excite modes with wavenumbers $k=\pm \omega^*$. In figure~\ref{fig:eigvalsfull} the growth rate predicted from the eigenvalues using the method of \citet{gupta1974theoretical} is shown as \added{a} function of the wavenumber. All types of modes excite stable and unstable modes (eigenvalues with positive growth $\sigma$ are the unstable modes). A heave motion with frequency $\omega^*$ excite eight eigenmodes:
\begin{itemize}
  \item Two unstable eigenmodes with $k=\omega^*-1$ and $k=\omega^*+1$ and their complex conjugates ($k=-(\omega^*-1)$ and $k=-(\omega^*+1)$);
  \item Two stable eigenmodes with $k=\omega^*-1$ and $k=\omega^*+1$ and their complex conjugates ($k=-(\omega^*-1)$ and $k=-(\omega^*+1)$);
\end{itemize}
while the surge motion excites four eigenmodes: stable and unstable modes with $k=\pm\omega^*$. As can be seen in equation~(\ref{eq:distz_comp}), the sway motion excites the same modes of the heave motion, only with a phase difference of $\pi/2$. The example of heave and sway motions for $\omega^*=1.5$ is shown as an example in figure~\ref{fig:eigvalsfull}. The modes with $k=\omega^*-1$ and $k=\omega^*+1$ are referred as the primary modes and the modes $k=-(\omega^*-1)$ and $k=-(\omega^*+1)$ as their complex conjugate.

\begin{figure*}[ht]
    \centering
    (a) \includegraphics[width=0.7\textwidth]{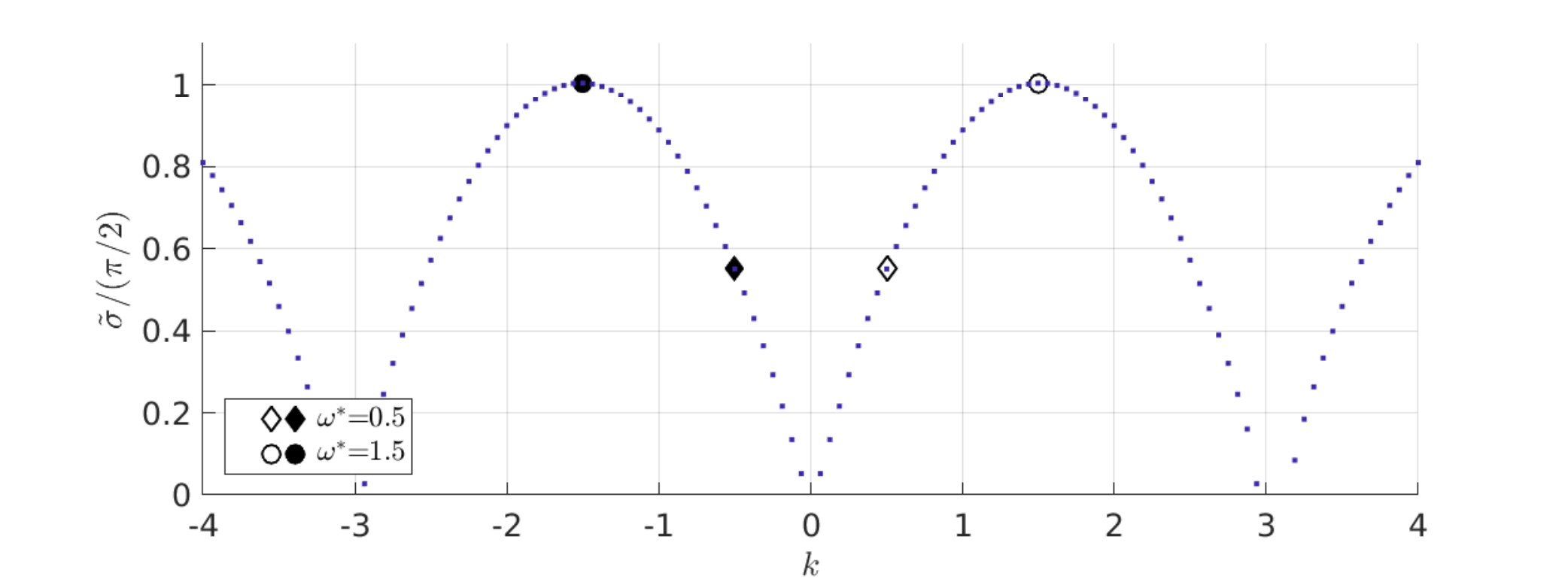} \\
    (b) \includegraphics[width=0.7\textwidth]{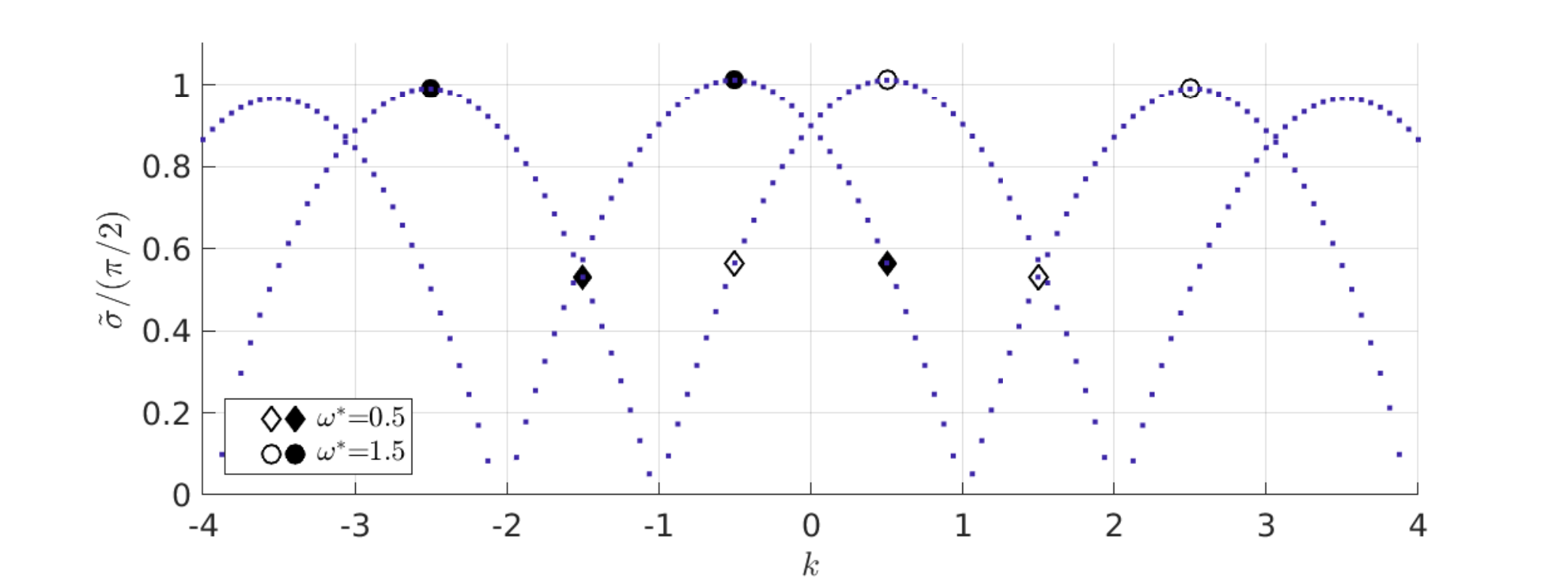} \\
    \caption{Theoretical growth rate for the unstable eigenvalues for a triple helix, split between the modes with phase correlation between the vortices (a) and modes with phase difference (b). Large symbols correspond to modes relevant to case $\omega^*=0.5$ and $\omega^*=1.5$.
    Filled symbols correspond to the complex conjugate of the primary mode. Surge modes are present in figure (a). Heave and sway modes are present in figure (b).}
    \label{fig:eigvalssplit}
\end{figure*}

The stability diagram of figures~\ref{fig:eigvalsfull} and \ref{fig:eigvalssplit} is usually only shown in the literature for positive values of $k$, since it is symmetric. However, since the primary mode $k=\omega^*-1 = -0.5$ for $\omega^*=0.5$ falls on the negative side, it is useful to look at the two-sided diagram. In the same sense, to say that a real (without imaginary part) perturbation excites wavenumber $k=\omega^*-1$ is equivalent to saying that the perturbation excites the conjugate pair with wavenumber $k=\pm(\omega^*-1)$.

Even knowing the wavenumber of a perturbation is not sufficient to know which unstable mode is excited. For most wavenumbers, there are three corresponding unstable eigenvalues, as can be seen in figure~\ref{fig:eigvalsfull} (for higher wavenumbers or higher values of pitch, this may be different, see examples in Ref.~\onlinecite{quaranta2019local}).

Some of the eigenvalues predicted by the stability theory correspond to modes that have a phase correlation between all vortices at $z=0$ while others have a phase difference of $\pm 2 \pi/N_b$ (see \citet{sarmast2014mutual,quaranta2019local} for complementary discussion about these modes). This is due to the symmetry of the stability problem regarding the $N_b$ helical vortices in the azimuthal direction. From the surge motion term of equation~(\ref{eq:distz_comp}), we note that there is no phase shift between the vortices (at the same streamwise position, the amplitude of the perturbation is the same for all helices). Hence, the eigenmodes that have a phase correlation are those excited by the surge motion, shown in figure~\ref{fig:eigvalssplit}(a).

The heave and sway motion have terms with a phase shift of exactly $\pm \phi_n = 2 \pi n/N_b$ between the vortices, see equation~(\ref{eq:distz_comp}), thus they excite the modes with a phase difference, shown in figure~\ref{fig:eigvalssplit}(b). In that  figure, there are two eigenvalues for each wavenumber. This can be explained by the two possible values of the phase difference between the vortices. One of the branches has a phase difference $2 \pi/N_b$ and the other a phase difference of $- 2 \pi/N_b$ (the branch without phase difference was already extracted from the figure and shown in figure~\ref{fig:eigvalssplit}(a)). Looking at equation~(\ref{eq:distz_comp}), it is expected that each primary mode excited by one frequency belongs to a different branch, because one term has phase difference $+ \phi_n$ and the other $- \phi_n$. This is exactly what is observed: the mode with $k=\omega^*-1$ belongs to one branch (related to $- \phi_n$) and the mode with $k=\omega^*+1$ belongs to the other branch (related to $+ \phi_n$). It is easy to see that the complex conjugate of the primary mode belongs to the opposite branch.

\begin{figure}[ht]
  \centering
  \includegraphics{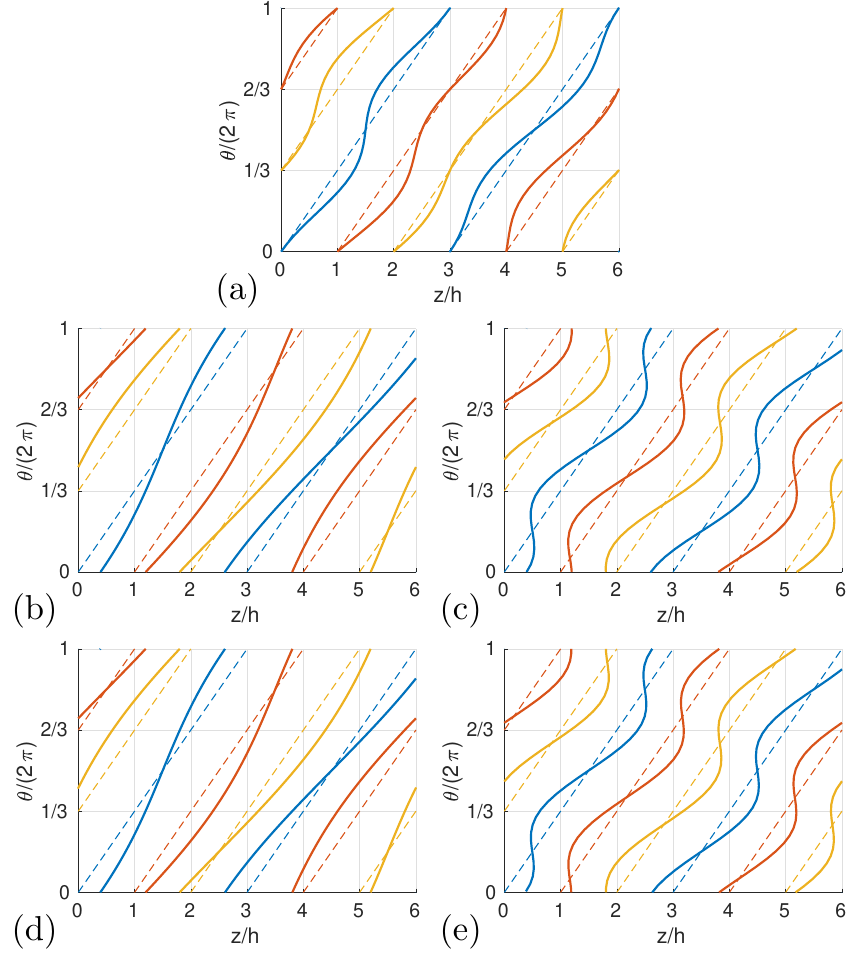}
  \caption{Comparison of the perturbation imposed by heave to the eigenvectors of the stability theory. Unrolled tip vortices, indicating distinct vortices by distinct colors. Dashed line: undisturbed helix. (a) Perturbation on helix for $\omega^*=0.5$ ($y$-component of equation \ref{eq:distz}). (b) and (c) Perturbation on helix decomposed in components with $k=-0.5$ and $k=1.5$, respectively ($y$-components of equation \ref{eq:distz_comp}). (d) Sum of eigenvectors of stability theory of helical vortices corresponding to eigenvalues indicated in figure~\ref{fig:eigvalssplit}(b) for $k=\pm 0.5$, obtained using the method of \citet{gupta1974theoretical}. (e) Sum of eigenvectors corresponding to $k=\pm 1.5$.}
  \label{fig:disturb05}
\end{figure}

\begin{figure}[ht]
  \centering
  \includegraphics{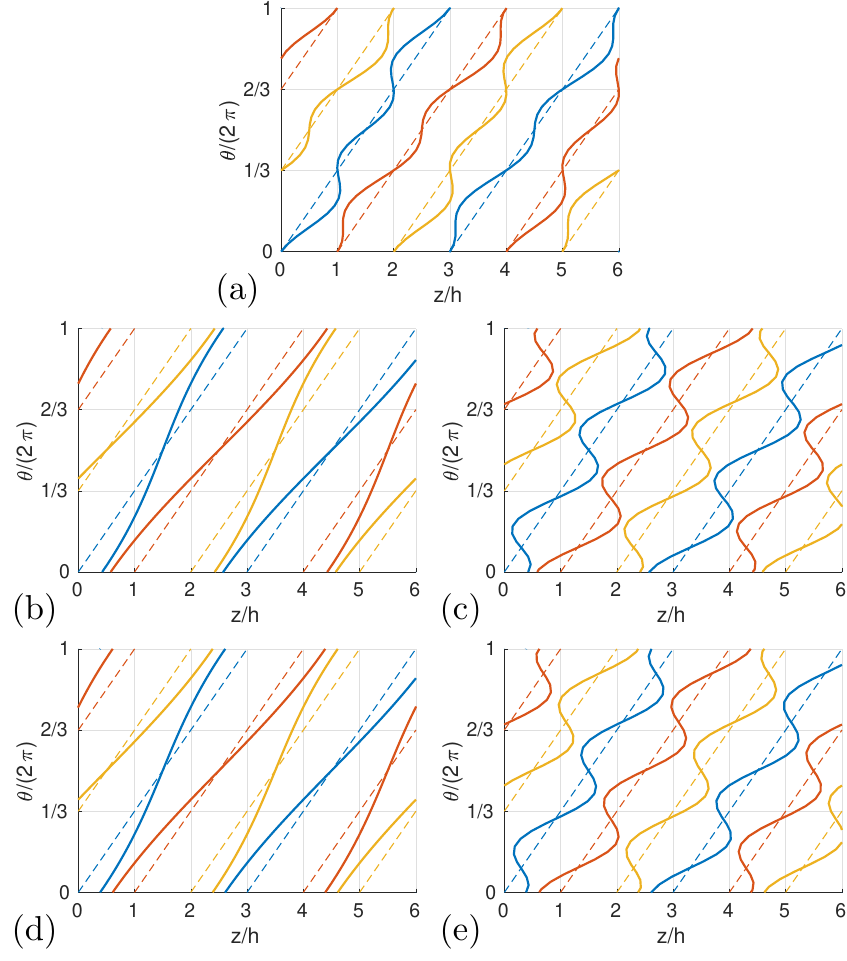}
  \caption{Comparison of the perturbation imposed by heave to the eigenvectors of the stability theory. Unrolled tip vortices, indicating distinct vortices by distinct colors. Dashed line: undisturbed helix. (a) Perturbation on helix for $\omega^*=1.5$ ($y$-component of equation \ref{eq:distz}). (b) and (c) Perturbation on helix decomposed in components with $k=0.5$ and $k=2.5$, respectively ($y$-components of equation \ref{eq:distz_comp}). (d) Sum of eigenvectors of stability theory of helical vortices corresponding to eigenvalues indicated in figure~\ref{fig:eigvalssplit}(b) for $k=\pm 0.5$, obtained using the method of \citet{gupta1974theoretical}. (e) Sum of eigenvectors corresponding to $k=\pm 2.5$.}
  \label{fig:disturb15}
\end{figure}

As can be seen in figures~\ref{fig:disturb05} and  \ref{fig:disturb15}, the perturbations created by the heave motion (equation~(\ref{eq:distz_comp})) are almost exactly the same as the eigenmodes predicted by the stability theory of \citet{gupta1974theoretical}. In order to obtain real eigenvectors, each complex eigenvector was summed with its complex conjugate, applying the adequate phase. Even though the figures were obtained with completely different methods, they look identical. For all other cases, a similar agreement was found. Perturbations for sway motion can also be reconstructed from the same eigenmodes, using appropriate values of phase. For surge motion (not shown here) the same agreement was observed when comparing to the modes of figure~\ref{fig:eigvalssplit}(a), when considering the modes with $k=\pm\omega^*$. This indicates that the perturbations imposed by the turbine motion agree almost identically to eigenvectors of the stability theory, thus, only these modes would be excited.

In Ref.~\onlinecite{kleine2021stability} we provided a rule-of-thumb to identify which eigenvalue corresponds to each heave frequency when there are multiple eigenvalues for the same wavenumber. The eigenvector that is observed to be identical to the perturbation is that whose primary eigenvalue in the parabolas of figure~\ref{fig:eigvalssplit}(b) has the same relative position in the parabola of 2-d row of vortices (figure~\ref{fig:eigvalsrow}). For example, for $\omega^*=0.5$:
\begin{itemize}
    \item There are two eigenvalues for $k=\omega^*-1=-0.5$, one close to $\tilde{\sigma}/(\pi/2) \approx 0.55$ and another close to $\tilde{\sigma}/(\pi/2) \approx 1.0$. The eigenvector that is observed to be identical to the perturbation (see figure~\ref{fig:disturb05}) is that whose eigenvalue has growth rate $\tilde{\sigma}/(\pi/2) \approx 0.55$, because its position in relation to the parabola between $-1<k<2$ is the same of $\omega^*=0.5$ in figure~\ref{fig:eigvalsrow} (diamond-shaped symbol). (The two eigenmodes for the pair $k=\pm 0.5$ with phase shift are shown in figures~\ref{fig:disturb05}(d) and \ref{fig:disturb15}(d), while the mode without phase shift is equivalent to figure~\ref{fig:Fmodes15}(e)).
    \item There are two eigenvalues for $k=\omega^*+1=1.5$, both close to $\tilde{\sigma}/(\pi/2) \approx 0.55$ (it is close to the intersection of parabolas between $-1<k<2$ and between $1<k<4$). The eigenvector that is observed to be identical to the perturbation (see figure~\ref{fig:disturb05}) is that whose eigenvalue belongs to parabola between $1<k<4$, because the position of the eigenvalue in relation to this parabola is the same of $\omega^*=0.5$ in figure~\ref{fig:eigvalsrow}.
\end{itemize}
The term ``parabola'' was freely used here, but we note that the curves do not represent exact parabolas, except for figure~\ref{fig:eigvalsrow}, that comes from equation~(\ref{eq:growth2d}). The discussion presented here provides an explanation for this rule-of-thumb: the branch that corresponds to $- \phi_n$ is similar to the branch without phase difference (figure~\ref{fig:eigvalssplit}(a)), but displaced one unit to the negative direction in the $k$-axis ($k=\omega^*-1$), while the branch that corresponds to $+ \phi_n$ is similar to the branch without phase difference, but displaced one unit to the positive direction in the $k$-axis ($k=\omega^*+1$).

Due to this connection between the excited modes and the stability of 2-d row of vortices, the results of stability of the full three-dimensional helical system might not be needed to estimate the growth rate of perturbations imposed by turbine motion. It is expected that the growth rate can be predicted from the theory of infinite row of two-dimensional vortices, within the restrictions discussed in Refs.~\onlinecite{quaranta2015long,quaranta2019local}.

The surge motion $\omega^*=1.5$ corresponds to the vortex pairing mechanism that has the highest growth rate $\tilde{\sigma} \approx \pi/2$, which has been discussed in previous studies that excite the vortices using body forces \citep{ivanell2010stability,sarmast2014mutual,kleusberg2019tip}. Additionally, we note that all motions with an angular frequency $\omega^*=1.5$ would also excite the vortex pairing mechanism that has the highest growth rate, even if the wavenumbers are $k = 0.5$ and $k = 2.5$. Figure~\ref{fig:disturb15} shows how the wavenumbers $k = 0.5$ and $k = 2.5$ correspond to the vortex pairing, in which neighboring vortices are out-of-phase in relation to each other. The mechanism for which the vortex pairing is excited by wavenumbers different than $k=N_b(j+1/2)$ (where $j$ is an integer) has been discussed in Refs.~\onlinecite{sarmast2014mutual,quaranta2019local,brynjell2020numerical}.

We emphasize here, however, that the only frequencies of motion of a turbine that excite the vortex pairing with $\tilde{\sigma} \approx \pi/2$ are given by $\omega^*=N_b(j+1/2)$, even if other wavenumbers are excited. This is valid not only for surge and heave but for all types of motion where the turbine moves as one. The fact that the turbine moves as one restricts the phase between helices to only two options: either there is no phase shift and $k = \omega^*$ (\emph{e.g.} surge) or the phase shift is $\pm 2\pi/3$ and $k = \omega^* \pm 1$ (\emph{e.g.} heave). This means, for example, that a frequency of $\omega^*=0.5$ would never excite the highest growth mode $\tilde{\sigma} \approx \pi/2$ for $k=0.5$, seen in figure~\ref{fig:eigvalssplit}(b). If the excitation is given by other forms, such as actuation on the blades or inflow turbulence, this restriction does not exist and vortex pairing may be induced by other frequencies.

The excitation of wavenumbers different from the normalized angular frequency for heave and sway is due to the terms $\sin \theta$ and $\cos \theta$ in equation~(\ref{eq:perturbations}). These terms can be thought of as coming from the decomposition of the oscillation of $y$ (for heave) or $z$ (for sway) in the radial direction, creating the phase shift between the perturbations on each vortex. Because surge does not have such terms, the wavenumber is equal to the normalized angular frequency. Hence, roll motion, equation~(\ref{eq:perturbationsroll}), would also create wavenumbers given by $k=\pm(\omega^*\pm1)$. Pitch, equation~(\ref{eq:perturbationspithc}), and yaw, equation~(\ref{eq:perturbationsyaw}), have components with and without terms $\sin \theta$ and $\cos \theta$. Therefore these motions would induce perturbations composed of several wavenumbers: the terms $R A_p$ and $R A_{yw}$ induce wavenumbers $k=\pm(\omega^*-1)$ and $k=\pm(\omega^*+1)$ and the terms $R_p A_p$ and $R_{yw} A_{yw}$ induce wavenumber $k=\pm\omega^*$.

A limitation of this model is the approximation that the perturbations are only advected in the streamwise direction. If the wake rotation is relevant, then the perturbations are also advected in the azimuthal direction, which would modify the wavenumbers induced by a certain frequency. Also, the model only considers mutual inductance and long-wave instabilities, which can be calculated from the theory of \citet{gupta1974theoretical}. For very large wavenumbers the long-wave assumption is not valid~\citep{quaranta2015long}. Short-wave instabilities \citep{kerswell2002elliptical,fabre2004short,fukumoto2005curvature,hattori2009short} are not captured by the model.

\section{Results} \label{sec:results}

\subsection{Direct numerical simulations} \label{sec:num_results}

Several cases with different types of motion and parameters were simulated, according to table~\ref{tab:cases}. The reference case has amplitude $A/R=1\%$ and angular frequency $\omega=0.5\Omega$, according to estimates detailed in section~\ref{sec:turbinemotion}. The amplitude of pitch was defined so that the maximum displacement in $z$ at the center of the turbine is $1\%$ of the radius, assuming the linear approximation ($1\%R/R_p=0.667\%$). The effect of the increase in amplitude is to advance the onset of the instability, bringing it closer to the turbine, as discussed in Ref.~\onlinecite{kleine2021stability}, similar to the increase in the amplitude of body forces or turbulence level shown in previous works~\citep{ivanell2010stability,sarmast2014mutual,kleusberg2019tip}. Since the effect of the amplitude is well understood, in this work we focus on the effect of the frequency and type of motion. A different amplitude, of $A/R=0.1\%$, is only used in section~\ref{sec:stab_results}, in order to compare the growth rate to analytical stability predictions.

\begin{table}[ht]
    \caption{Parameters of motion for the cases shown in figures~\ref{fig:vz_yz_vort} and \ref{fig:vz_xz_vort}.}
    \label{tab:cases}
    \centering
    \begin{ruledtabular}
    \begin{tabular}{c c c}
      Angular Frequency & Type of  & Amplitude \\
      $\omega^*=\omega/\Omega$ & motion & $A/R$ \\
      \hline
      $0.5$ & Heave\footnotemark[1] and surge\footnotemark[1] & $1\%$ \\
      $1$ & Heave & $1\%$\\
      $1.5$ & Heave\footnotemark[1] and surge\footnotemark[1] & $1\%$ \\
      $2.5$ & Heave & $1\%$ \\
      $0.5$ & Pitch & $0.667\%$ \\
    \end{tabular}
    \end{ruledtabular}
    \footnotetext[1]{Cases chosen to be quantitatively compared to stability analysis, with an amplitude of $A/R=0.1\%$, in section~\ref{sec:stab_results}.}
\end{table}

\begin{figure*}[tp]
  \centering
  \includegraphics[scale=0.99]{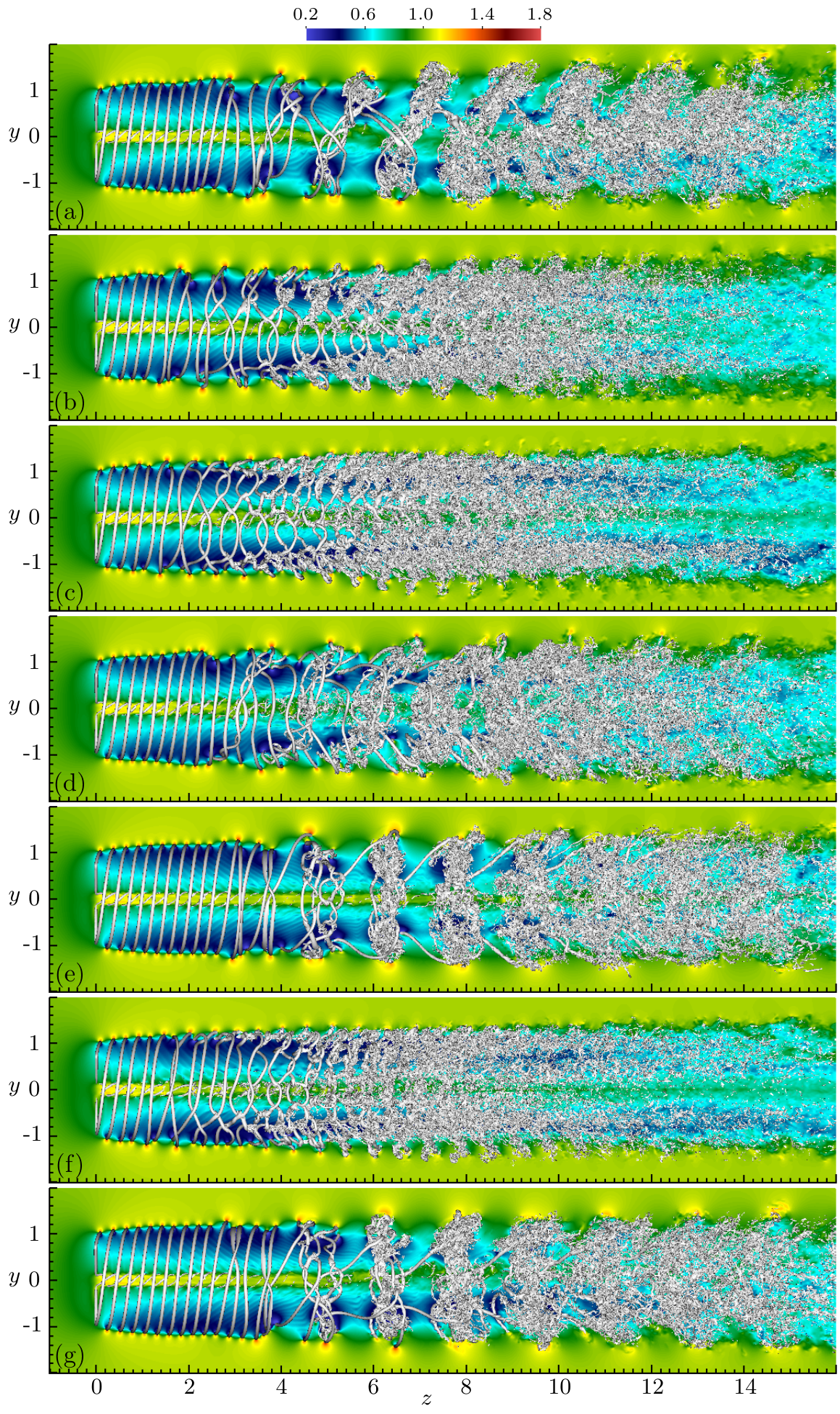}
  \caption{Instantaneous streamwise velocity along the wake in the $yz$-plane. Three dimensional iso-surfaces of vorticity magnitude ($|\xi| = 15$) are shown in grey. (a) Heave  $\omega^*=0.5$. (b)  Heave  $\omega^*=1.0$. (c) Heave  $\omega^*=1.5$. (d) Heave  $\omega^*=2.5$. (e) Surge  $\omega^*=0.5$. (f) Surge  $\omega^*=1.5$. (g) Pitch  $\omega^*=0.5$.}
  \label{fig:vz_yz_vort}
\end{figure*}

\begin{figure*}[tp]
  \centering
  \includegraphics[scale=0.99]{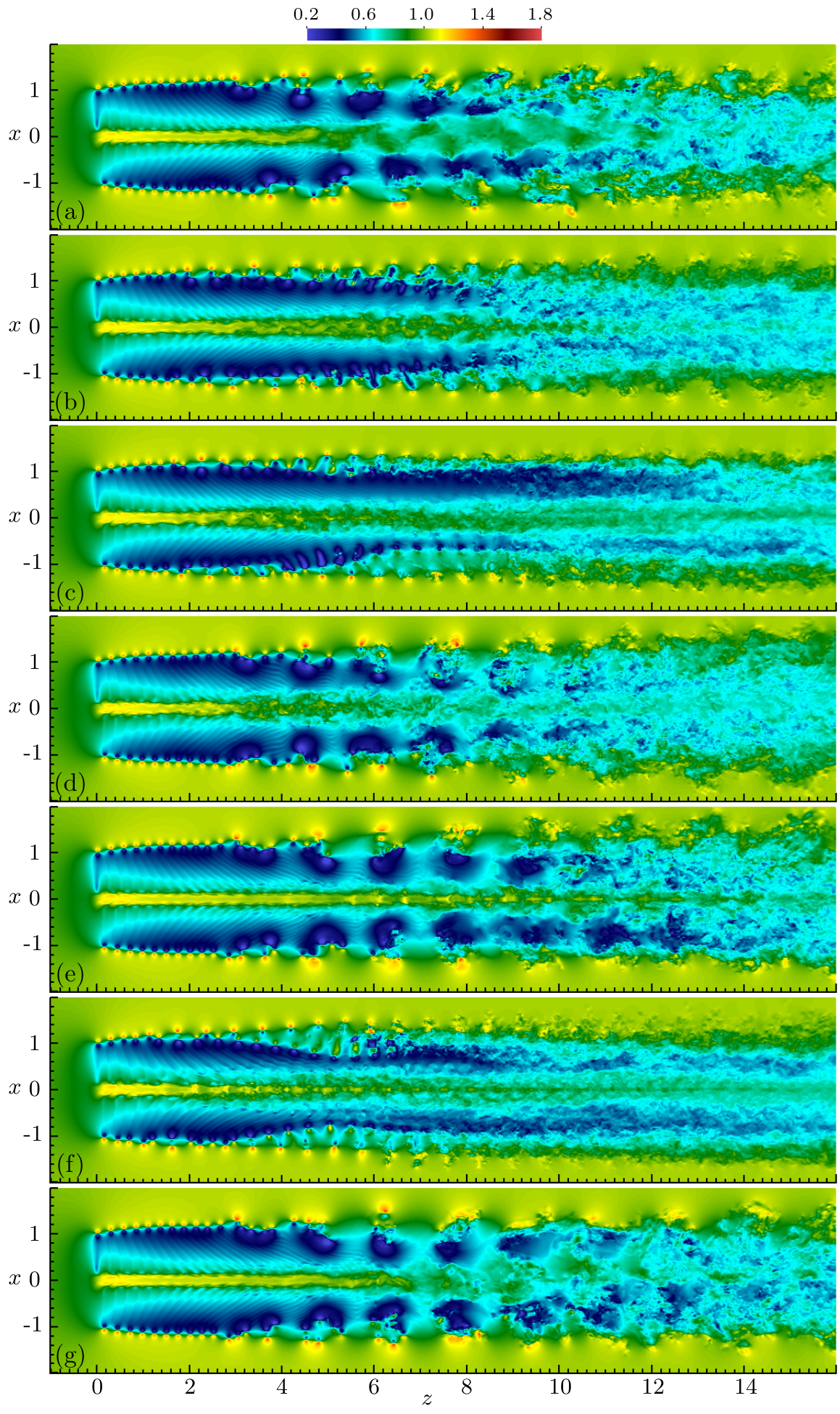}
  \caption{Instantaneous streamwise velocity along the wake in the $xz$-plane. (a) Heave  $\omega^*=0.5$. (b)  Heave  $\omega^*=1.0$. (c) Heave  $\omega^*=1.5$. (d) Heave  $\omega^*=2.5$. (e) Surge  $\omega^*=0.5$. (f) Surge  $\omega^*=1.5$. (g) Pitch  $\omega^*=0.5$.}
  \label{fig:vz_xz_vort}
\end{figure*}

The change in the frequency of motion highlights a noteworthy effect. For the lower frequency, $\omega^*=0.5$, larger flow structures are created inside the wake, as shown in figures~\ref{fig:vz_yz_vort} and \ref{fig:vz_xz_vort}. Observing the vortex interaction in figure~\ref{fig:vz_yz_vort}, it is possible to notice that these flow structures are created by the merging of the vortices. The formation of similar structures for lower frequencies can also be noted in the simulations reported in Ref.~\onlinecite{lee2019effects}. However, in their case, the lower frequency was accompanied by a modification in the nature of the motion, from translation (heave, sway and surge) to rotation (yaw, pitch and roll), making it difficult to isolate the effects. We show here that the existence of flow structures is independent of the type of motion, occurring for heave, surge and pitch (figures~\ref{fig:vz_yz_vort} and \ref{fig:vz_xz_vort}(a), (e) and (g)).

Interestingly, large flow structures, very similar to those observed for $\omega^*=0.5$, can also be noted for the higher frequency, $\omega^*=2.5$. According to the stability theory for a row of vortices, this is expected. Every frequency that can be written as $\omega^*-j N_b$, with $j$ integer, is indistinguishable from $\omega^*$ according to the theory presented in section~\ref{sec:stability2d}. It follows that the configuration of a row of vortices for $\omega^*=2.5$ is equal to $\omega^*=-0.5$, which is the mode complex conjugate of $\omega^*=0.5$. Thus, the number of vortices that interact directly is predicted to be the same for $\omega^*=2.5$ and $\omega^*=0.5$.

The number of vortices that interact with each other can be calculated from the maximum of $1/p$ and $1/(1-p)$ from the relationship $p=\omega^*/N_b-j$. The mechanism for $\omega^*=0.5$ is illustrated in figure~\ref{fig:vortexrow}: the vortices $-1 \le m \le 1$ would move to the right and the vortices $2 \le m \le 4$ would move to the left, forming a group of $6$ vortices that interact directly with each other. The values of table \ref{tab:interac_vort} agree well with figure~\ref{fig:vz_yz_vort}. For example, regardless of the type of motion, in the numerical simulations for $\omega^*=0.5$, six vortices interact with each other and coalesce to form the large flow structure discussed in section~\ref{sec:num_results}. For $\omega^*=1.5$, two vortices interact with each other and the vortex pairing mechanism is excited, as expected. The type of motion does not influence the number of vortices that interact, as long as the normalized frequency is the same, as predicted by the theory. Even the case of $\omega^*=2.5$, which has a more complex dynamics, seems to follow this rule. For this case the number of vortices that interact is difficult to count, however, the distance between the large flow structures is approximately $6h$, as predicted.

\begin{table}[ht]
    \caption{Number of interacting vortices as function of angular frequency predicted by the theory of section~\ref{sec:stability2d}.}
    \label{tab:interac_vort}
    \centering
    \begin{ruledtabular}
    \begin{tabular}{ c  c  c  c  c}
      $\omega^*$ & $p=\frac{\omega^*}{N_b}$ & $\frac{1}{p}$ & $\frac{1}{1-p}$ & Number of interacting vortices \\ 
      \hline
      0.5 & 1/6 & 6 & 6/5 & 6  \\
      1.0 & 1/3 & 3 & 3/2 & 3\\
      1.5 & 1/2 & 2 & 2 & 2 \\
      2.5 & 5/6 & 6/5 & 6 & 6 \\
    \end{tabular}
    \end{ruledtabular}
\end{table}

The shape of the large flow structures is influenced by the type of motion. For $\omega^*=0.5$, the surge motion creates flow structures that are coherent in the azimuthal direction while the heave creates flow structures that are tilted. The explanation for this difference is discussed in section~\ref{sec:stab_results}. The pitch motion is more similar to surge, which is expected since $R_p > R$. Nonetheless, the structures created by pitch are not as coherent in the azimuthal direction as those of surge, showing an effect of the $(\sin \theta)$ term, similar to heave.

The streamwise flow inside the wake experiences higher fluctuations due to these flow structures. These fluctuations might affect downstream turbines, increasing unsteady loads that may worsen the fatigue. In the case of a downstream FOWT, it may increase the amplitude of its motion. It is even possible to imagine an extreme case, where the unsteady flow and the sea waves add to each other with the same frequency, creating very high amplitudes. It is especially worth investigating because the frequency of the flow structures created for $\omega^*=0.5$ is the same frequency as the upstream heave motion, caused by sea waves.

In order to estimate the effect of the unsteadiness of the streamwise velocity on a downstream turbine, the parameters $F_z^*$ and $T_y^*$ were defined based on integrals of the streamwise dynamic pressure. $F_z^*$ is an indicator of the streamwise force
that a fictitious disk, with the same radius of the turbine ($R=1$) centered at $(x_c,y_c,z_c)$, would be subjected to. It is the streamwise dynamic pressure integrated over the area of the disk, normalized by the equivalent value at infinity
\begin{equation}
    F_z^*=\frac{\int_S \frac{1}{2} \rho w^2 dS}{\frac{1}{2} \rho W_{\infty}^2 S},
    \label{eq:dynpress}
\end{equation}
where $S$ is the area of the disk and $\rho$ is the fluid density. $T_y^*$ is an indicator of the torque around the negative $y$ direction (torque around the tower, that tends to yaw a turbine), defined as
\begin{equation}
    T_y^*=\frac{\int_S \frac{1}{2} \rho w^2 (x-x_c) dS}{\frac{1}{2} \rho W_{\infty}^2 \frac{1}{2} S \frac{4}{3\pi} R},
    \label{eq:dynpressT}
\end{equation}
where the normalization is performed considering an extreme case where the velocity on the semicircle of $x-x_c<0$ is zero and the velocity on the semicircle of $x-x_c>0$ (area $S/2$ and centroid at $4R/(3\pi)$) is constant and equal to $W_{\infty}$.

\begin{figure*}[ht]
    \includegraphics[width=1.0\textwidth]{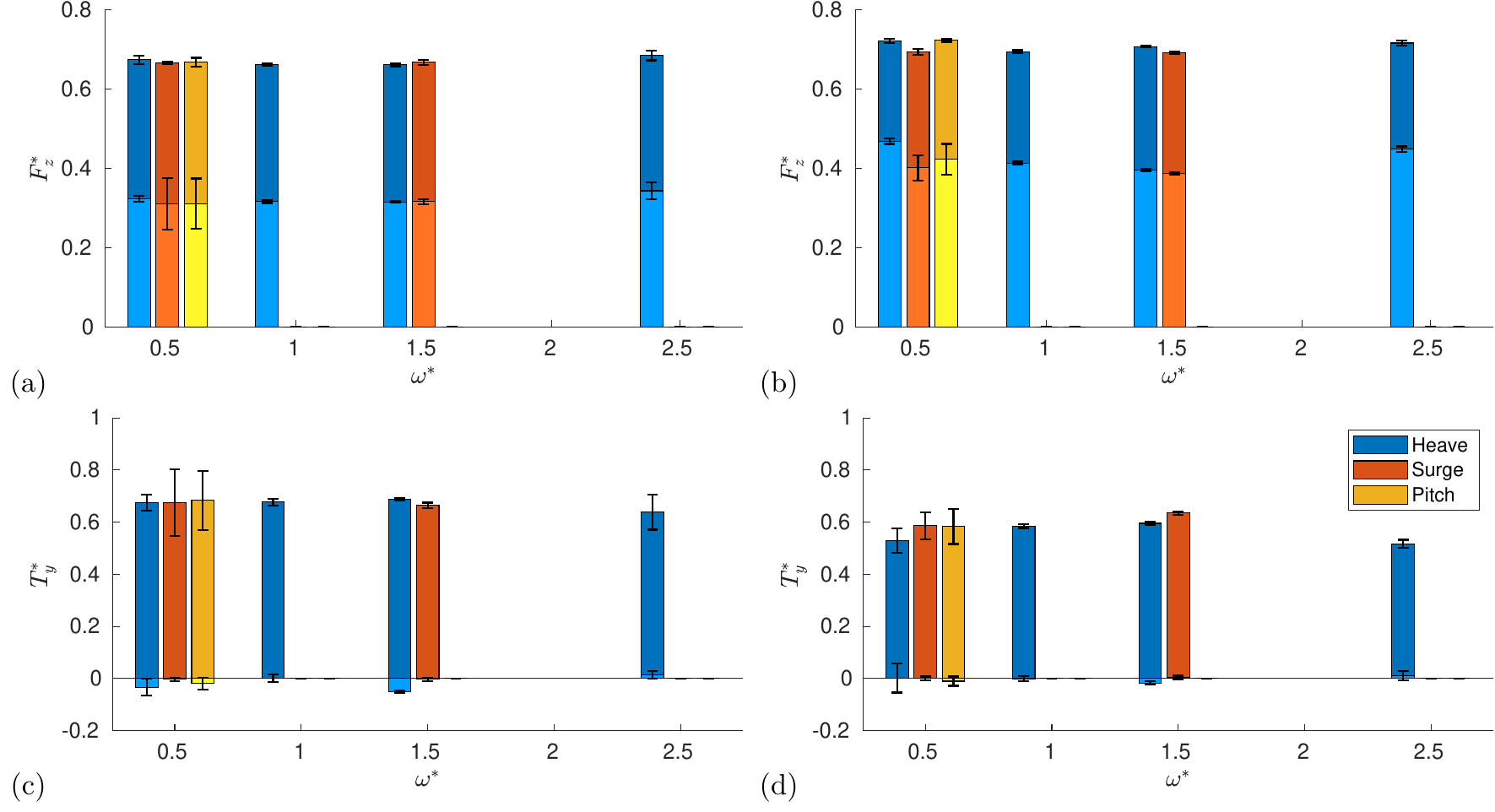}
    \caption{Average in time of $F_z^*$ and $T_y^*$, for a disk completely inside the wake (in light color, $(x_c,y_c)=(0,0)$) and a disk only partially in the wake (in darker color, $(x_c,y_c)=(1,0)$). Error bars indicate root mean square (RMS) of $F_z^*$. (a, c) $z_c=6$. (b, d) $z_c=12$.}
    \label{fig:dynpress}
\end{figure*}

Four positions for the disk were chosen for this calculation. The disks were placed at the streamwise positions $z_c=6$ and $z_c=12$, in positions completely inside the wake, $(x_c,y_c,z_c)=(0,0,6)$ and $(x_c,y_c,z_c)=(0,0,12)$, and positions partially inside the wake, $(x_c,y_c,z_c)=(1,0,6)$ and $(x_c,y_c,z_c)=(1,0,12)$. The values of the average in time of $F_z^*$ and $T_y^*$ are shown in figure~\ref{fig:dynpress}, with error bars indicating their root mean square (RMS). In table \ref{tab:mean_rms_dynpress}, the ratio of RMS to the average of $F_z^*$ is shown. The values of table \ref{tab:mean_rms_dynpress} deviates from the values found in Ref.~\onlinecite{kleine2021stability} because there was a programming error in the code that calculated the RMS of that work.

\begin{table}[ht]
    \caption{Oscillation in time of the integral of the dynamic pressure, $F_z^*$ (RMS divided by average).}
    \label{tab:mean_rms_dynpress}
    \centering
    \begin{ruledtabular}
    \begin{tabular}{c |c c c c|c c|c}
      Disk Position & \multicolumn{4}{c|}{Heave ($\omega/\Omega$)} & \multicolumn{2}{c|}{Surge ($\omega/\Omega$)} & \multicolumn{1}{c}{Pitch ($\omega/\Omega$)} \\
      $(x_c,y_c,z_c)$  & 0.5 & 1 & 1.5 & 2.5 & 0.5 & 1.5 & 0.5\\ 
      \hline
      (0,0,6) & 2.3\% & 1.1\% & 0.5\% & 6.1\% & 21.0\% & 2.1\% & 20.6\% \\
      %\hline
      (1,0,6) & 1.6\% & 0.4\% & 0.6\% & 1.8\% & 0.5\% & 1.0\% & 1.6\% \\
      %\hline
      (0,0,12) & 1.5\% & 0.8\% & 0.8\% & 1.5\% & 8.0\% & 1.0\% & 9.1\% \\
      %\hline
      (1,0,12) & 0.8\% & 0.5\% & 0.3\% & 0.9\% & 1.1\% & 0.4\% & 0.5\% \\
    \end{tabular}
    \end{ruledtabular}
\end{table}

As can be observed in table \ref{tab:mean_rms_dynpress} and figure~\ref{fig:dynpress}, in general, the RMS of the integrals of the dynamic pressure are higher for angular frequencies $\omega^*=0.5$ and $\omega^*=2.5$, both for heave and surge. This indicates that these frequencies might be more dangerous for downstream turbines. As expected, a great oscillation in in $T_y^*$ is observed for a disk partially in the wake ($(x_c,y_c)=(1,0)$). Surge and pitch cause great oscillation in $F_z^*$ for a turbine completely inside the wake, because the flow structures are coherent in the azimuthal direction. This would cause high amplitudes in the forces applied to the tower and mooring cables of a downstream turbine. If these effects occur in real-world situations, it might be necessary to employ motion damping, wake steering, change to the rotation frequency or other control mechanisms to avoid a turbine completely inside the wake of an upstream turbine under surge or pitch with one of the dangerous frequencies. Nevertheless, the effect of a sheared inflow alone has been shown to break the coherence in the azimuthal direction \citep{kleusberg2019tip}, which might suggest that the most worrying values of table \ref{tab:mean_rms_dynpress} have low probability of occurring in a real situation.\added{ For a sheared inflow, without inflow turbulence, \citet{kleusberg2019tip} showed that the breakdown of the vortices in the wake occurs first in the positions with lower streamwise velocity, which usually correspond to the region closer to the ground (i.e., with lower $y$ values). Hence, even if the disturbance is azimuthally coherent, the flow structures will not be azimuthally coherent if the inflow is not uniform.}

In general, the oscillations are reduced with the distance from the turbine, which can be explained by the intrinsic turbulent mechanisms of the wake breaking the structures of concentrated vorticity, as can be seen in figures~\ref{fig:vz_yz_vort} and \ref{fig:vz_xz_vort}. Without inflow turbulence, the large flow structures are still present at a distance of 6 diameters from the turbine and cause relevant velocity fluctuations. Hence, the effects of turbine motion can demand higher distances between the turbines. However, before any conclusion is drawn, these phenomena need to be better understood in more realistic inflow conditions, especially with incoming turbulence and taking into consideration structural properties and the dynamic effects of motion on the operation conditions for the turbine.

\added{The Reynolds number is not believed to have a relevant role on the initial growth of the instabilities and on the first non-linear effects, since the main mechanisms of vortex interaction are inviscid, as described in section~\ref{sec:stability}. Also, the vortex core size, which is a function of the Reynolds number and of the parameters of the actuator line model (as discussed in section~\ref{sec:turbinemodel}, see also Refs.~\onlinecite{dag2020new,martinez2019filtered}), has only a second-order effect on the stability and dynamics of the vortices in the near wake. However, the Reynolds number and vortex core properties may affect the vortex breakdown and mixing inside the wake, which has an effect on the dynamic pressure, especially for the most downstream position. Hence, there are reasons to believe that the qualitative aspects of table~\ref{tab:mean_rms_dynpress} are valid for different Reynolds numbers, but more studies are needed to understand its influence.}

As expected, the average value of $F_z^*$ is lower for the disks closer to the turbine ($z_c=6$) and inside the wake, while the average value of $T_y^*$ higher for the disks closer to the turbine and is almost zero inside the wake. Another effect that can be seen for heave in figures~\ref{fig:dynpress}(b, d) is that $F_z^*$ inside the wake is higher and $T_y^*$ partially in the wake is lower for $\omega^*=0.5$ and $\omega^*=2.5$ when compared to the other frequencies. One possible explanation is that the large flow structures provide better mixing with the external flow, bringing streamwise momentum to the wake. This is an indicator that the large flow structures might also have a beneficial effect. The creation of large, coherent flow structures was identified by \citet{yilmaz2018optimal} as a possible flow control strategy of wind farms. They showed that the creation of coherent vortex rings in the wake of a turbine is an optimum way to increase the power production of a configuration of two in-line wind turbines without inflow turbulence. Turbine motion can be further explored as a possible strategy to create such coherent structures that increase power production on wind farms.\added{ It should be noted that this strategy is different from the proposal of destabilizing the tip vortices in order to accelerate the wake recovery (e.g., Refs.~\onlinecite{huang2019numerical,marten2020predicting,brown2022accelerated}), where a more uniform wake is obtained.}

\subsection{Comparison to the stability theory} \label{sec:stab_results}

In figure~\ref{fig:vz_yz_vort}, it is possible to note that the interaction of vortices occurs further upstream for frequencies closer to $\omega^*=1.5$, while the region of interaction for frequencies $\omega^*=0.5$ and $\omega^*=2.5$ seems to be comparable. This agrees qualitatively with the stability theory (section~\ref{sec:stability}), which predicts the highest growth rate for $\omega^*=1.5$ for all types of motion. In order to quantitatively compare the growth rate of perturbations with the stability theory, a lower amplitude of motion of $0.1\%$ is used in the numerical simulations. An amplitude of $1\%$ is too close to the saturation amplitude where non-linear effects start to be relevant, making the linear region too short to allow calculation of the linear growth. The shorter domain is used ($z_{out}=11.78$), since we are interested in the near-wake dynamics. Qualitatively, the flow behavior is similar to the simulations with an amplitude of $1\%$, but the transition to turbulence occurs further upstream, as can be seen in Ref.~\onlinecite{kleine2021stability}.

\begin{figure}[ht]
    \centering
    (a)
    \includegraphics[width=0.45\textwidth]{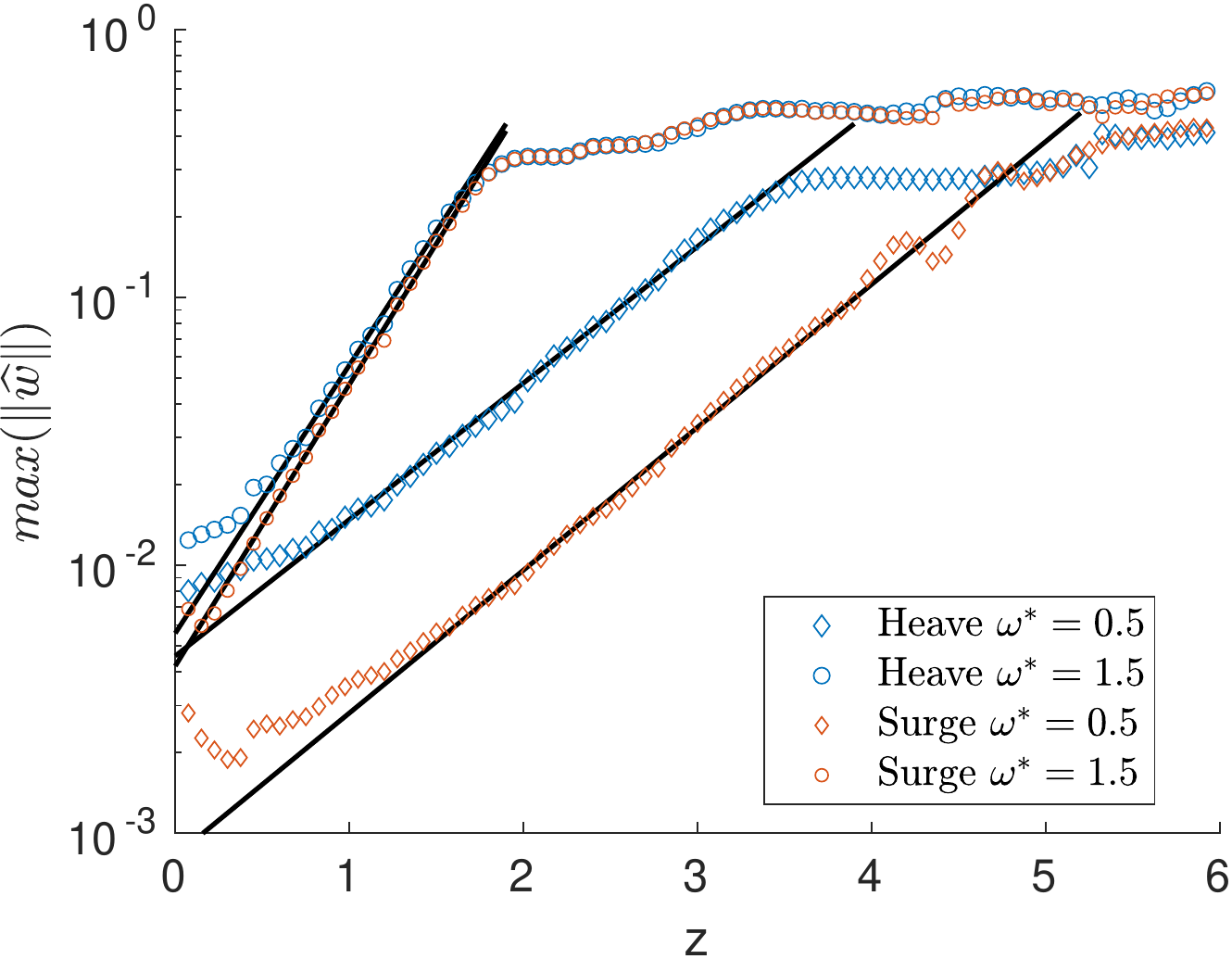} \\
    (b)
    \includegraphics[width=0.45\textwidth]{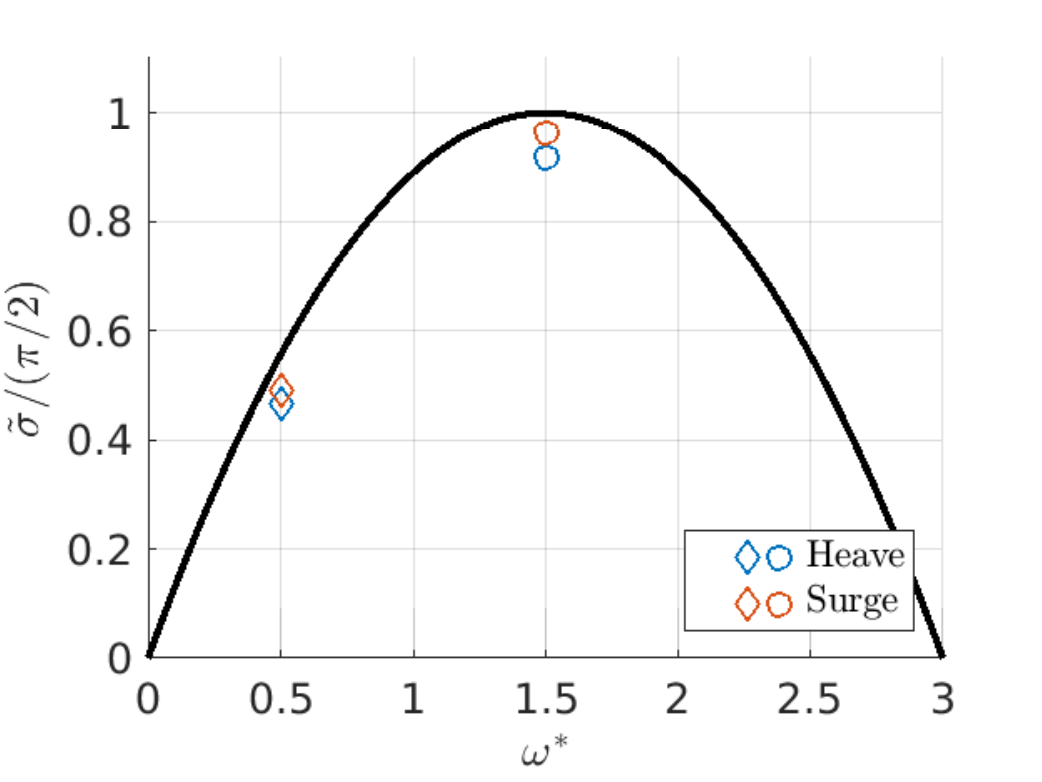}
    \caption{(a) Growth of each perturbation in the streamwise direction. Black lines correspond to exponential fit. (b) Growth rate from the simulations compared to the value predicted by stability theory of a row of vortices.}
    \label{fig:growth_wz}
\end{figure}

In order to quantitatively compare the spatial growth rate of the perturbations, the flow is decomposed in its Fourier series in time, as performed in Refs.~\onlinecite{ivanell2010stability,kleusberg2019tip}. During a period $\Delta T$, $N_S=48$ flow-field snapshots $\phi_q$, $q=0,\dots,N_S-1$, equally spaced in time are used to extract the component $\widehat{\Phi}$ in each frequency,
\begin{equation}
    \widehat{\Phi}_{n_f}=\frac{1}{N_S}\sum^{N_S-1}_{q=0} \phi_q e^{-i q n_f \frac{2\pi}{N_S}},
\end{equation}
where $i=\sqrt{-1}$ and $n_f$ is the index (integer) that defines the selected angular frequency, given by $\omega=n_f 2 \pi/\Delta T$. By selecting the frequency of motion for each simulation, the growth of the perturbations can be obtained by analysing the evolution in space of the maximum of the absolute value of $\widehat{\Phi}$ \citep{ivanell2010stability,kleusberg2019tip}. The Fourier mode of the streamwise component of the velocity, $\widehat{w}$, is used to calculate the growth rate $\sigma = d(\log(max(\widehat{w})/W_\infty))/dz$, which is then nondimensionalized and scaled according to $\tilde{\sigma}=\sigma (2h^2 w_c/\Gamma)$ \citep{sarmast2014mutual,kleusberg2019tip}. \added{Similarly to other studies~\citep{ivanell2010stability,kleusberg2019tip,quaranta2019local}, after a region of receptivity, the growth is exponential until non-linear effects become relevant and the growth saturates (figure~\ref{fig:growth_wz}(a)). In order to calculate the growth rate, only the region where the growth could be considered exponential was considered.}

As can be seen in figure~\ref{fig:growth_wz}\added{(b)}, the growth rate is very close to the values predicted by the stability theory. However, the growth rates are slightly lower, especially for perturbations imposed by heave. Previous works \citep{kleusberg2019tip} showed better agreement when using local properties to scale the growth rate of perturbations imposed as body forces to the tip vortices. Since the turbine motion modifies the wake as a whole, more than just perturbing the tip vortices, a possible hypothesis is that these more complex dynamics also influence the stability of the wake. The higher difference for heave seems to agree with this hypothesis, because heave motion also modifies the shape of the wake, while the boundary of the near wake is undisturbed for surge. More studies are needed to understand this difference. Nevertheless, the good agreement indicates the instabilities predicted by the linear theory are the dominant effect.

\begin{figure*}[ht]
  \centering
  \includegraphics{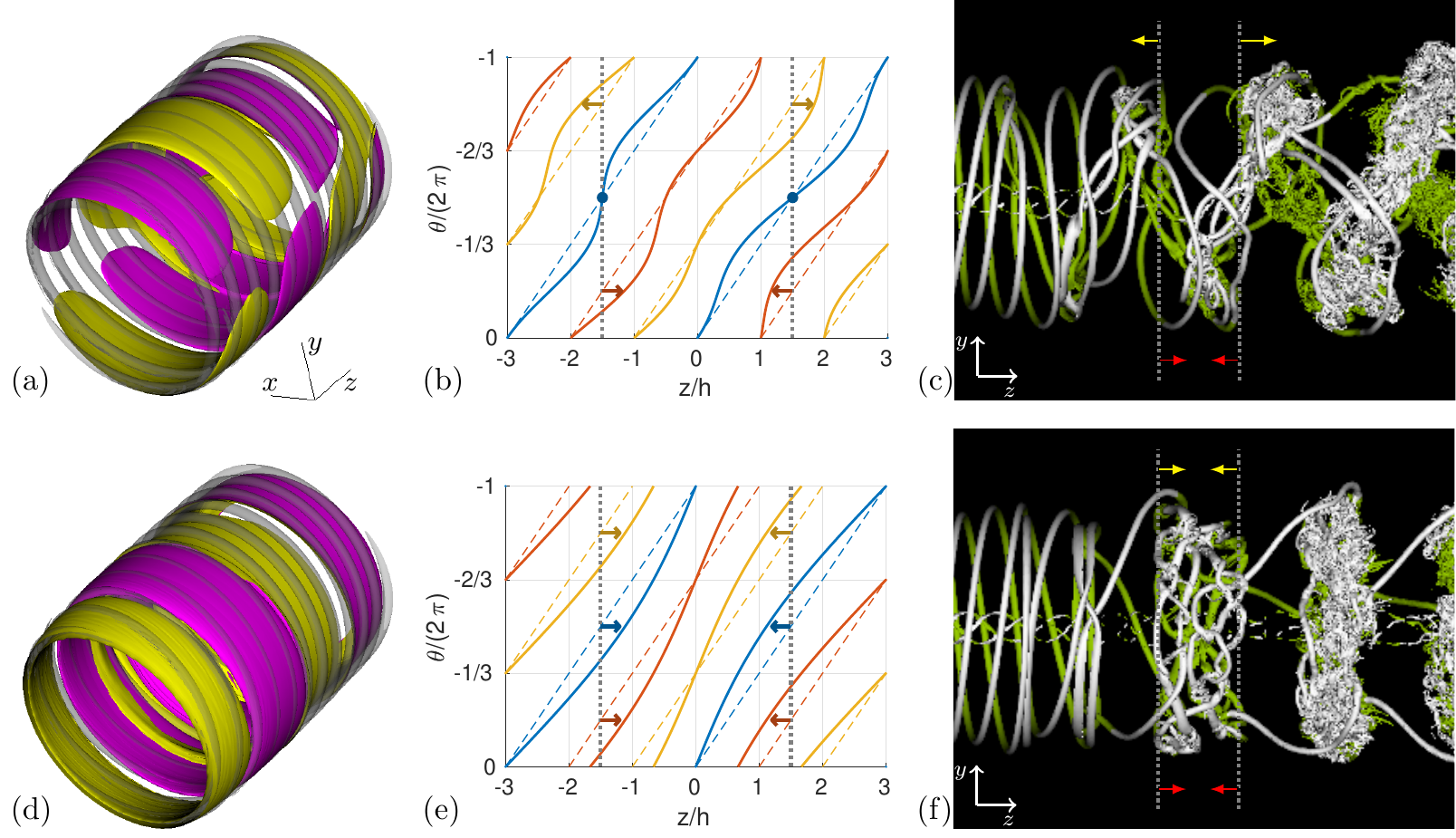}
  \caption{(a, d) Real part of the Fourier mode $\widehat{w}$ for simulations with $A/R=0.1\%$ and $\omega^*=0.5$, depicted using positive (purple) and negative (yellow) contours for heave (top row) and surge (lower row), respectively (contours were cropped to show only the values close to the region of the tip vortices, therefore, the growth is not visible). (b, e) Perturbation predicted by the first-order approximation (note inverted ordinate axis). (c, f) Highlight of the vortex interaction of simulations with $A/R=1\%$, with arrows that indicate the direction in which the vortices coalesce. To indicate perspective, the contours of vorticity in the back ($x<0$) are indicated in green, while contours of vorticity in the front ($x>0$) are indicated in white.}
  \label{fig:Fmodes05}
\end{figure*}

\begin{figure*}[ht]
  \centering
  \includegraphics{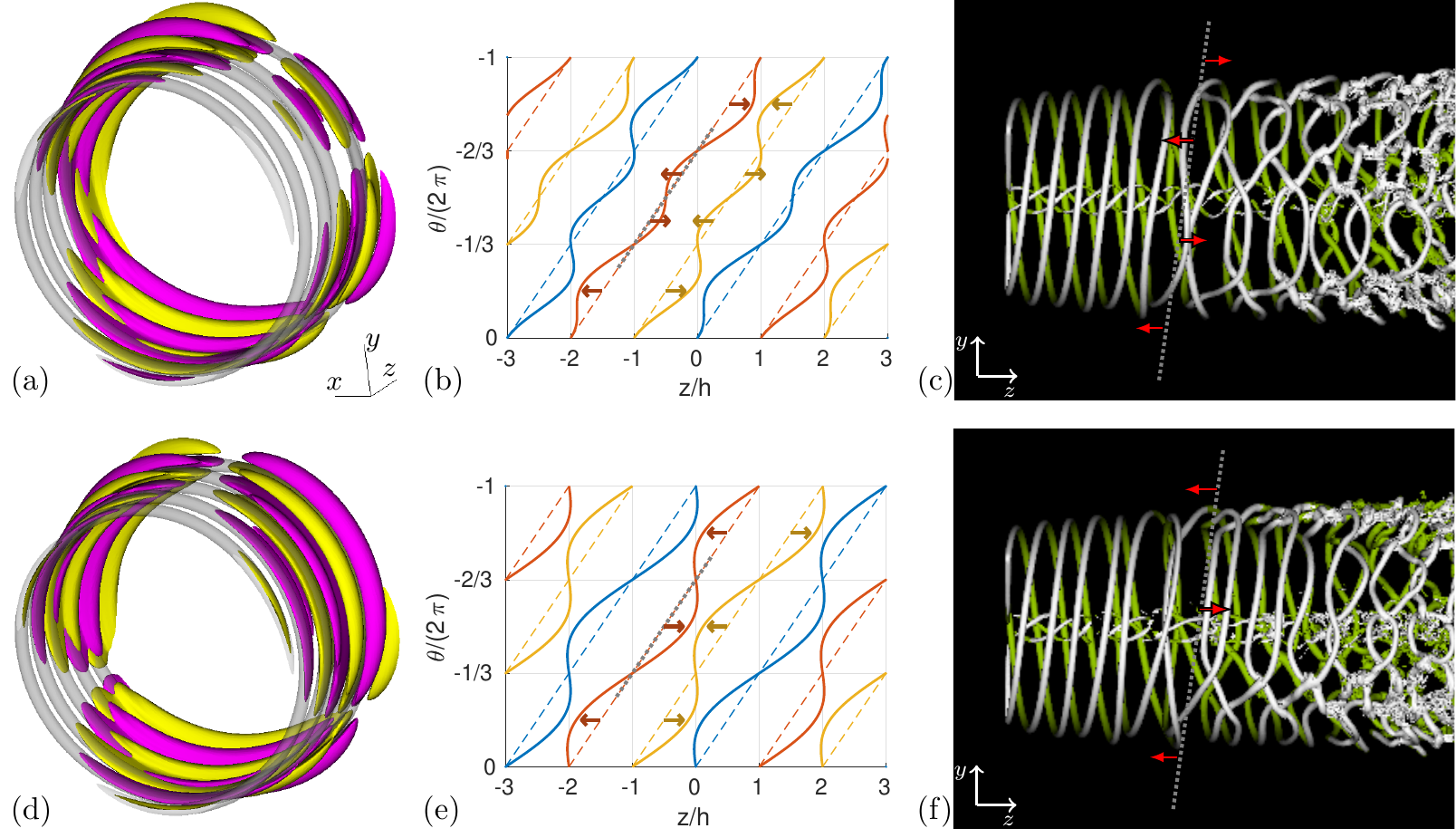}
  \caption{(a, d) Real part of the Fourier mode $\widehat{w}$ for simulations with $A/R=0.1\%$ and $\omega^*=1.5$, depicted using positive (purple) and negative (yellow) contours for heave (top row) and surge (lower row), respectively. (b, e) Perturbation predicted by the first-order approximation (note inverted ordinate axis). (c, f) Highlight of the vortex interactions of simulations with $A/R=1\%$, with arrows that indicate the direction in which a disturbance in a segment of vortex (in white) tends to grow. To indicate perspective, the contours of vorticity in the back ($x<0$) are indicated in green, while contours of vorticity in the front ($x>0$) are indicated in white.}
  \label{fig:Fmodes15}
\end{figure*}

Figures~\ref{fig:Fmodes05} and \ref{fig:Fmodes15} show the real part of the Fourier modes. The perturbations imposed by the heave and surge motions, from section~\ref{sec:perturbationshs}, are also shown. The perturbations (shown in figures~\ref{fig:Fmodes05} and \ref{fig:Fmodes15}(b, e)) were adapted to the left-handed helices, following the reference of equation~(\ref{eq:vortposref1}) (instead of equation~(\ref{eq:vortposref}), used in most of section~\ref{sec:stability}).

The similarities between the first-order approximation of the perturbations and the Fourier modes that are present in the wake are clear, \replaced{which}{what} indicates that the modes predicted in section~\ref{sec:stability} dominate the flow in the near wake. Also, even for the simulations with an amplitude of motion of $1\%$, where the linear region is short, the stability theory and the first-order perturbation can explain the flow structures that are seen in the non-linear region of the flow. In figure~\ref{fig:Fmodes05}, looking at two positions $z$ indicated by gray dashed lines, we note that the vortices with positive $y$ are displaced in opposite direction to vortices with negative $y$ for heave, while for surge all vortices at a certain position $z$ are displaced in the same direction. This explains why the flow structures for heave are tilted. The results for pitch motion seen in figure~\ref{fig:vz_yz_vort}(g) show a combination of the effects: the vortices coalesce in one large structure coherent in $z$, but the vortices in the negative $y$ position coalesce faster than the vortices with positive $y$, since it is formed by the sum of the effects of surge and heave, weighted by the ratio $R_p/R$, as discussed in section~\ref{sec:perturbationpwr}.

In figure~\ref{fig:Fmodes15}, following one vortex, indicated by a gray dashed line, we note that between $-\pi/3 > \theta > -2\pi/3$ the heave motion has two lobes while the surge motion only has one lobe. This effect can be noted in the Fourier mode of the simulation with low amplitude (a,d), the perturbation (b,e) and in the displacement of the vortex in the non-linear region of the simulation with higher amplitude (c,d).

Based on the results presented here, we believe that the mechanism that forms the near-wake structures can then be explained as follows. The perturbation on the tip vortices, imposed by the turbine motion, excites vortex instability modes which are practically identical to the perturbation itself. These modes grow exponentially in time and space until non-linear effects become significant. The flow structures that are formed in the non-linear regime are a consequence of this mechanism and preserves some of the characteristics predicted by the linear theory.

Only one type of motion and frequency is considered in each numerical simulation presented here. The linear theory, by assumption, treats different types of motion and frequencies independently. The good agreement of our numerical results with the linear theory suggests that, if there is more than one type of motion or frequencies involved, the one that will dominate is that with the higher growth rate according to the linear stability theory. However, further studies are needed to verify if non-linear effects might be relevant for certain cases.

As can be seen in figures~\ref{fig:vz_yz_vort} and \ref{fig:vz_xz_vort}, the type of motion influences the flow near the center of the turbine. The flow near the hub vortices seems much less disturbed for surge (figure (e)) than for heave (figure (a)). The linear stability of helical vortices might suggest an explanation. The analysis of section~\ref{sec:stability2d} is not applicable for the hub vortices because the approximations of low pitch are not valid. The radial position of the hub vortices is lower than the distance between neighboring vortices. In this case, a more elaborate model is needed. First, we note that the approximations that $\delta z \approx \delta r$ and the azimuthal component is not relevant are not valid. For higher pitch, a perturbation in the streamwise direction has lower importance (in the limit of infinite pitch, the vortex is indifferent to a perturbation in the streamwise direction). Hence, the perturbations imposed by heave are more similar to the eigenvectors of the stability theory than perturbations imposed by surge. Besides that, \citet{quaranta2019local} showed, by transposing the results of \citet{robinson1982three} to the helical system, that the growth rate of the modes in each ``parabola'' of the growth diagram decreases as the wavenumber increases, an effect that is more pronounced for higher pitch (this effect that can also be seen in figure~\ref{fig:eigvalsfull}, to a lesser extent). The ``parabola'' with the highest growth is that with the peak closer to $k=0$, which corresponds to modes with a phase difference (figure~\ref{fig:eigvalssplit}(b)), that are excited by the heave motion. Thus, even if the modes were directly excited, the mode excited by surge would have a lower growth rate. Therefore, both effects indicate that perturbations in the hub vortices by heave motion will reach higher amplitude earlier. In real turbines, the nacelle and its wake will greatly influence the flow in this region, so it is uncertain if such effect would occur in more realistic configurations. Further analyses of the effect of motion on the hub vortices are left for future studies.

All the analyses performed in this work considered uniform inflow without turbulence. In the case of sheared inflow or yawed turbine, tip vortex instabilities from the linear stability theory were already shown to dominate the near-wake stability when vortices are disturbed by body forces \citep{kleusberg2019tip,kleusberg2019wind}. Hence, we believe that the mechanism described in this study for the formation of the near wake of turbines under motion would also be applicable to sheared inflow conditions or yawed turbines, with the necessary adaptations to the theory of section~\ref{sec:stability}.

However, the modes described in this work might not dominate the flow for a FOWT in a turbulent atmospheric boundary layer. The turbulence level found in real-world applications might be in the same order of magnitude or higher than the perturbations imposed by the motion. Thus, free-stream turbulence could dominate the near wake, preventing the growth of the perturbations imposed by turbine motion. The large flow structures created by turbine motion with $\omega^*=0.5$ only occur due to the phase shift between different helices being restricted because the turbine moves as one. Inflow turbulence does not have this restriction, and, therefore, several frequencies could have growth rate $\tilde{\sigma} \approx \pi/2$, higher than the growth of disturbances with $\omega^*=0.5$, as demonstrated by \citet{sarmast2014mutual}. Alternatively, turbulence could dissipate or modify flow structures created in the near wake in such a way that they might not be relevant for a downstream turbine. High turbulence level has been shown to accelerate the disintegration of coherent vortex rings in the simulations of \citet{yilmaz2018optimal}. Also, for a fixed-bottom wind turbine with turbulent inflow, \citet{decillis2022influence} observed that the modes due to tip vortex instabilities are superseded by modes related to flow fluctuations coming from incoming turbulence. Nevertheless, the motion of floating turbines imposes very specific perturbations, that are not present on fixed-bottom turbines. Hence, further studies are needed to investigate the effects of turbulence on moving turbines.

\section{Conclusions} \label{sec:conclusions} %% \conclusions[modified heading if necessary]

The wake of a floating offshore wind turbine was analysed by means of numerical simulations and comparison with \deleted{the} linear stability theory. For the numerical simulations, the use of actuator lines with a high-order SEM code made the analysis of the vortex interactions possible. Two simplified models of vortex stability that are applicable to all types of motion were developed: a model based on the stability of a row of vortices and a model based on the stability of helical vortices.

Using a first-order approximation, it was noted that the perturbations induced by the heave and sway motion have wavenumbers of $k=\omega^*-1$ and $k=\omega^*+1$, in contrast to the perturbations imposed by surge forces, which have wavenumber $k=\omega^*$. Modes compatible with these wavenumbers were observed in the numerical simulations in the near-wake and formed flow structures that are observed in the non-linear region of the wake.

Despite the different wavenumbers created by different types of motion, the growth rate depends mostly on the frequency for helices of low pitch and is compatible with a model of an infinite row of two-dimensional vortices calculated by \citet{lamb1932hydrodynamics}. The stability theory of a row of two-dimensional vortices also predicted the number of vortices that interact with each other in the numerical simulations. For $\omega^*=1.5$, our simulations showed the out-of-phase mechanism of two vortex pairing observed in other numerical \citep{ivanell2010stability, sarmast2014mutual,kleusberg2019tip} and experimental works \citep{felli2011mechanisms,quaranta2015long,quaranta2019local}. For $\omega^*=0.5$ and $\omega^*=2.5$, larger structures with the interaction of six vortices were observed for all types of motion simulated. These larger flow structures created higher oscillations of the streamwise velocity inside the wake. The RMS of the integrated dynamic pressure was considerably higher for this case, especially for surge and pitch motions, which create coherent structures in the azimuthal direction. This could increase the fatigue loads and the amplitude of the motion of downstream FOWTs.

Our simulations, without inflow turbulence, showed that turbine motion excites the unstable modes of helical vortices. These modes grow in time and space, forming unsteady flow structures that may affect downstream turbines. Free-stream turbulence, not considered here, could dissipate or modify these flow structures. Hence, future studies are necessary to confirm if these effects are sustained under more realistic conditions.

The results suggest that the linear theory can be used to predict which modes imposed by turbine motion will dominate the flow if more than one type of motion or frequency is present. However, further studies are needed to understand the impact of non-linear interactions. This study shows that the general qualitative behaviour of the wake can be understood and predicted from relatively simple stability models. A good understanding of the wake effects of turbines under motion is necessary to develop distancing parameters for wind farms of FOWTs. Knowledge about which frequencies and types of motion influence downstream turbines can also guide control laws that avoid dangerous situations or promote favorable flow conditions. %?

\begin{acknowledgements}
The computations were performed on resources provided by the Swedish
National Infrastructure for Computing (SNIC) at the PDC Center for High-Performance Computing at the
Royal Institute of Technology (KTH) and the National Supercomputer Centre at Link{\"o}ping University. This work was conducted within StandUp for Wind. VGK thanks KTH Engineering Mechanics for partially funding this work. BSC acknowledges the support from FAPESP (Funda{\c c}{\~a}o de Amparo {\`a} Pesquisa do Estado de S{\~a}o Paulo), Proc. 2019/01507-8, for this research. BSC also thanks the Brazilian National Council for Scientific and Technological Development (CNPq) for financial support in the form of a productivity grant, number 314221/2021-2.
\end{acknowledgements}

\section*{AUTHOR DECLARATIONS}

This article may be downloaded for personal use only. Any other use requires prior permission of the author and AIP Publishing. This article appeared in Physics of Fluids (2022) and may be found at \url{https://doi.org/10.1063/5.0092267}.

\subsection*{Conflict of Interest}

The authors have no conflicts to disclose.

\section*{Data Availability}

The data that support the findings of this study are available from the corresponding author upon reasonable request.

\appendix 
\section{First-order approximation of perturbation imposed by
yaw and roll} \label{app:perturbationyr}    %% Appendix A

To derive equation~(\ref{eq:bladepitch}), it is assumed that the center of rotation for pitch is located on the $y$-axis of the turbine. The same assumption is applied here for roll. The center of rotation for yaw is assumed on the $x$-axis of the turbine. For most configurations of FOWT, these are reasonable assumptions. Nevertheless, the formulas presented here would not be applicable for unusual configurations in which the distance in $z$-direction between the center of rotation and the turbine is in the same order of magnitude of $R$ or the distance in other directions.

Yaw motion (subscript $_{yw}$) is conceptually similar to pitch, hence it also generates similar components 
\begin{equation}
  \begin{aligned} &
    \begin{bmatrix}
      \delta r_v  \\
      \delta z_v 
    \end{bmatrix} = \\ &
    \begin{bmatrix}
      0 \\
      R_{yw} A_{yw} \sin{(\omega_{yw}^*( \theta-\phi_n))} - R A_{yw} \sin{(\omega_{yw}^*( \theta-\phi_n))} \cos \theta 
    \end{bmatrix}
  \end{aligned}
    \label{eq:perturbationsyaw}
\end{equation}
where the position of the center of rotation, $R_{yw}$, would usually be zero, since the axis of rotation is normally the tower of the turbine. For unusual configurations, such as platforms with multiple turbines, $R_{yw}$ can be different from zero.

Analogously, for roll motion (subscript $_r$),
\begin{equation}
  \begin{aligned} &
    \begin{bmatrix}
      \delta x_v \\
      \delta y_v \\
      \delta z_v
    \end{bmatrix} = \\ &
    \begin{bmatrix}
      R_r A_r \sin{(\omega_r^*( \theta-\phi_n))} - R A_r \sin{(\omega_y^*( \theta-\phi_n))} \sin \theta  \\
      R A_r \sin{(\omega_y^*( \theta-\phi_n))} \cos \theta \\
      0
    \end{bmatrix}
  \end{aligned}
\end{equation}
which becomes
\begin{equation}
    \begin{bmatrix}
      \delta r_v  \\
      \delta z_v 
    \end{bmatrix} =
    \begin{bmatrix}
      R_r A_r \sin{(\omega_r^*( \theta-\phi_n))} \cos \theta \\
      0
    \end{bmatrix} ,
    \label{eq:perturbationsroll}
\end{equation}
where the component with amplitude $R A_r$ affects only the azimuthal perturbation while its radial component is cancelled out. Hence, for the purpose of our simplified analysis that neglects the azimuthal component, the roll motion is mainly similar to a sway motion of amplitude $R_r A_r$.

%% REFERENCES
\bibliography{references.bib}

\textcolor{blue}{\listofchanges}

\end{document}